\UseRawInputEncoding
\documentclass[11pt,aps,preprint]{revtex4}%
\usepackage{amsfonts}
\usepackage{amsmath}
\usepackage{amssymb}
\usepackage{multirow}
\usepackage{graphicx}%
\providecommand{\U}[1]{\protect\rule{.1in}{.1in}}

\begin{document}
\preprint{HEP/446421}
\title{Low-lying doubly heavy baryons: Regge relation and mass scaling }
\author{YongXin Song$^{1}$}
\author{Duojie Jia$^{1,2}$\thanks{Corresponding author}}
\email{jiadj@nwnu.edu.cn}
\author{WenXuan Zhang$^{1}$}
\author{Atsushi Hosaka$^{3}$}
\affiliation{$^{1}$Institute of Theoretical Physics, College of Physics and Electronic Engineering, Northwest Normal University, Lanzhou 730070, China }
\affiliation{$^{2}$Lanzhou Center for Theoretical Physics, Lanzhou University,
Lanzhou,730000,China }
\affiliation{$^{3}$Research Center for Nuclear Physics, Osaka University, 10-1 Mihogaoka, Ibaraki, Osaka 567-0047, Japan}

\begin{abstract}
In framework of heavy-diquark$-$light-quark endowed with heavy-pair binding, we explore excited baryons $QQ^{\prime}q$ containing two heavy quarks ($QQ^{\prime}=cc,bb,bc$) by combining the method of Regge trajectory with the perturbative correction due to the heavy-pair interaction in baryons. Two Regge relations, one linear and the other nonlinear, are constructed semi-classically in the QCD string picture, and a mass scaling relation based on the heavy diquark-heavy antiquark symmetry are employed between the doubly heavy baryons and heavy mesons. We employ the ground-state mass estimates compatible with the observed doubly charmed baryon $\Xi_{cc}$ and spectra of the heavy quarkonia to determine the trajectory parameters and the binding energies of the heavy pair, and thereby compute the low-lying masses of the excited baryons $\Xi_{QQ^{\prime}}$ and $\Omega_{QQ^{\prime}}$ up to 2S and 1P excitations of the light quark and heavy diquark. The level spacings of heavy diquark excitations are found to be smaller generally than that of the light-quark excitations, in according with the nature of adiabatic expansion between the heavy and light-quark dynamics.
\end{abstract}

\maketitle

\section{Introduction}

Being systems analogous to hydrogen-like atoms with notable level splitting, by which QED interaction is tested in minute detail, the doubly heavy (DH) baryon
provides a unique opportunity to probe the fundamental theory of the strong
interaction, quantum chromodynamics (QCD) and models inspired by
it \cite{PEMP,Mehen:D17,EFG:D02,KR:D14}. It is expected that the excited DH baryons can form a set of nontrivial levels with various mass splitting and confront the QCD interaction with the experiments in a straightforward way.

In 2017, the LHCb Collaboration at CERN discovered
the doubly charmed baryon $\Xi_{cc}^{++}$ in the $\Lambda_{c}^{+}K^{-}\pi
^{{+}}\pi^{+}$ mass spectrum \cite{AAIJ:17BC} and reconfirms it in the decay
channel of $\Xi_{c}^{+}\pi^{+}$ \cite{AAIJ:18BA}, with the measured mass
$3621.55\pm0.23\pm0.30$ MeV \cite{AAIJ:20J}. This observation sets up, for the
first time, a strength scale of interaction between two heavy quarks and is of
help to understand the strong interaction among heavy
quarks \cite{KR:D14,SotoCast:D21,Mehen:D17}. For instance, the mass of the $\Xi_{cc}^{++}$ helps to \textquotedblleft calibrate\textquotedblright\ the binding energy between a pair of heavy quark governed by short-distance interaction. Further more, very recent observation of the doubly charmed tetraquark \cite{TccLHCb:2021} is timely to examine the assumed interquark ($QQ$) interaction in hadrons and invite conversely systematic study of the DH baryons. DH baryons have been studied over the years using various theoretical methods such as the nonrelativistic (NR) \cite{Gershtein:2000nx,Kiselev:2002iy,Roberts:2007ni,KR:D14} as well as relativistic \cite{EFG:D02} quark models, heavy-quark effective theory \cite{KKP94}, QCD sum rules \cite{BCN93,ZH08,ZWang10}, the Feynman-Hellmann theorem \cite{RLP95}, mass formula \cite{BGH97}, the Skyrmion model \cite{RRS90}, the lattice QCD \cite{MLW02,LMW01,FMT03,LLOW10,ACCDG12,NPCS13,PEMP} and the effective field theory \cite{PS09,SC20,SotoCast:D21}. Till now, the features(masses, spin-parity and decay widths) of the DH baryons in their excited states remain to be explored.

In this work, we introduce an energy term of heavy-pair binding due to the perturbative dynamics into the Regge relations of heavy hadrons to explore excited doubly heavy baryons with the help of the observed data of the light and heavy mesons, and of the $\Xi_{cc}^{++}$ as well as the ground-state masses of the DH baryons compatible with the measured data. A nonlinear Regge relation for the heavy diquark excitations in DH baryons, endowed with heavy-pair binding is constructed and applied to compute low-lying mass spectra of the excited DH baryons $\Xi_{QQ^{\prime}}$ and $\Omega_{QQ^{\prime}}$ ($QQ^{\prime}=cc,bb,bc$), combining with the mass scaling that is tested successfully for the experimentally available mass splittings of the heavy baryons. An agreement with other calculations is achieved and a discussion is given for the excitation modes of the DH baryons in the QCD string picture.

We illustrate that the QCD string picture \cite{Nambu74}, when endowed with the two-body binding energy of the heavy quarks, gives an unified description of the DH baryons in which both aspects of the long-distance confinement and perturbative QCD are included, similar to that of the effective QCD string (EFT) in Ref. \cite{SotoCast:D21}. While our approach is semiclassical in nature, it treats quark (diquark) and string dynamically and full relativistically, in contrast with the static(nondynamic) approximation of the heavy quarks in DH baryons in Ref. \cite{SotoCast:D21} and of the lattice QCD simulation \cite{PEMP}.

This paper is organized as follows. In Sect. II, we apply linear and nonlinear Regge relation to formulate and estimate the spin-independent masses of the excited DH baryons in the 1S1p, 2S1s and 1P1s waves. In Sect. III, we formulate the spin-dependent forces and the ensuing mass splitting of the low-lying excited DH baryons in the scheme of the $jj$ coupling. In Sect. IV, we estimate the spin coupling parameters via the method of mass scaling these parameters and compute the excited spectra of the DH baryons. We end with summary and remarks in Sect. V.

\section{Spin-independent mass of excited DH baryons}

We use heavy-diquark$-$light-quark picture for the DH baryons $qQQ^{\prime}$($q=u,d$
and $s$) with heavy quarks $QQ^{\prime}=cc,bb$ or $bc$, in which the heavy
pair $QQ^{\prime}$ can be scalar(spin zero) diquark denoted by $[QQ^{\prime}]$
or axial-vector(spin-$1$) diquark denoted by $\{QQ^{\prime}\}$. The charm
diquark $cc$ and the bottom diquark $bb$ can only form axial-vector diquark,
$\{cc\}$ or $\{bb\}$, while the bottom-charm diquark $bc$ can form both,
$\{bc\}$ or $[bc]$. We use the notations $N_{D}L_{D}nl$ to label the quantum
numbers of the DH baryons, with the value of principal quantum number($N_{D}$)
of diquark, its orbital momentum($L_{D}$) denoted by a capital letter and the
principal quantum number($n$) for the excitations of light quark and its
orbital momentum($l$) by a lowercase letter.

The mass of a DH baryon $qQQ^{\prime}$ consists of the sum of two parts: $M=\bar{M}+\Delta M$, where $\bar{M}$ is the spin-independent mass and $\Delta M$ is the mass splitting due to the spin-dependent interaction. As the diquark $QQ^{\prime}$ is heavy, compared to
the light quark $q$, the heavy-light limit applies to the DH baryons
$qQQ^{\prime}$.

\subsection{Low-lying DH baryons with ground-state diquark}
For a DH baryon with heavy diquark $QQ^{\prime}$ in internal S wave, we employ, by analogy with Ref. \cite{JiaLH:19}, a linear Regge relation derived
from the QCD string model(Appendix A). Basic idea of this derivation is to view the baryon to be a $q-QQ^{\prime}$ system of a massive QCD string
with diquark $QQ^{\prime}$ at one end and $q$ at the other. Denoting by $\bar
{M}_{l}$ the mass of a DH baryon with orbital angular momentum $l$ of light quark , the Regge relation takes form \cite{JiaLH:19}(Appendix A)%
\begin{equation}
(\bar{M}_{l}-M_{QQ^{\prime}})^{2}=\pi al+\left[  m_{q}+M_{QQ^{\prime}}%
-\frac{m_{\text{bare}QQ^{\prime}}^{2}}{M_{QQ^{\prime}}}\right]  ^{2},
\label{regge0}%
\end{equation}
where $M_{QQ^{\prime}}$ is the effective mass of the S-wave heavy diquark,
$m_{q}$ the effective mass of the light quark $q$, $a$ stands for the tension
of the QCD string \cite{Nambu74}. Here,
$m_{\text{bare}QQ^{\prime}}$ is the bare mass of $QQ^{\prime}$, given
approximately by sum of the bare masses of each heavy quark: $m_{\text{bare
}QQ^{\prime}}$$=m_{\text{bare}Q}+m_{\text{bare}Q^{\prime}}$. Using the bare data $m_{\text{bare,}c}$$=1.275 \text{ GeV}$ $m_{\text{bare,}b}$$=4.18 \text{ GeV}$, one has numerically,
\begin{equation}
m_{\text{bare }cc}=2.55~\text{GeV},m_{\text{bare }bb}=8.36\text{GeV}%
,m_{\text{bare }bc}=5.455\text{GeV}. \label{mbare}%
\end{equation}

We set the values of $m_{q}$ in Eq. (\ref{regge0}) to be that in Table I,
which was previously determined in Ref. \cite{JiaLH:19} via matching the measured mass spectra of the excited singly heavy baryons and mesons.

\renewcommand{\tabcolsep}{0.08cm}
\renewcommand{\arraystretch}{1.0}
\begin{table}[tbh]
\caption{The effective masses (in GeV) of the quark masses and the string
tensions $a$ (in GeV$^{2}$) of the $D/D_{s}$ and $B/B_{s}$ mesons in Ref. \cite{JiaLH:19}. }%
\label{tab:Eff-mass}
\begin{tabular}
[c]{ccccccccc}\hline\hline
\text{Parameters} & \text{$M_{c}$} & \text{$M_{b}$} & \text{$m_{n}$} &
\text{$m_{s}$} & \text{$a({c\bar{n})}$} & \text{$a({b\bar{n})}$} &
\text{$a({c\bar{s})}$} & \text{$a({b\bar{s}})$}\\\hline
Input & $1.44$ & $4.48$ & $0.23$ & $0.328$ & $0.223$ & $0.275$ & $0.249$ &
$0.313$\\\hline\hline
\end{tabular}
\end{table}

To determine $M_{QQ^{\prime}}$ in Eq. (\ref{regge0}), we apply Eq.
(\ref{regge0}) to the ground state ($n=0=l$) to find \cite{JiaLH:19}%
\begin{align}
\bar{M}(1S1s)  &  =M_{QQ^{\prime}}+m_{q}+\frac{k_{QQ^{\prime}}^{2}%
}{M_{QQ^{\prime}}},\label{ReggS}\\
k_{QQ^{\prime}}  &  \equiv M_{QQ^{\prime}}v_{QQ^{\prime}}=M_{QQ^{\prime}%
}\left(  1-\frac{m_{\text{bare}QQ^{\prime}}^{2}}{M_{QQ^{\prime}}^{2}}\right)
^{1/2}, \label{mv}%
\end{align}
which agrees with the mass formula given by heavy quark
symmetry \cite{Manohar:D07}. As the experimental data are still lacking except
for the doubly charmed baryon $\Xi_{cc}^{++}$, we adopt the masses (Table II)
of DH baryons computed by Ref. \cite{EFG:D02}, which successfully predicts $M(\Xi_{cc},1/2^{+})=3620$ MeV, very close to the measured data
$3621.55$ MeV of the doubly charmed baryon $\Xi_{cc}^{++}$ \cite{AAIJ:17BC,AAIJ:20J}. \renewcommand{\tabcolsep}{0.20cm}
\renewcommand{\arraystretch}{1.0}
\begin{table}[tbh]
\caption{Ground state masses (GeV) and their spin averages $\bar{M}_{l=0}$ of the doubly heavy baryons in Ref. \cite{EFG:D02}.}%
\label{tab:Spin-average mass}
\begin{tabular}
[c]{cccccc}\hline\hline
\text{State} & \text{$J^{P}$} & \text{Baryons} & Content & \text{Mass} &
\text{$\bar{M}_{l=0}$}\\\hline
$1^{2}S_{1/2}$ & $\frac{1}{2}^{+}$ & $\Xi_{cc}$ & $n\{cc\}$ & $3.620$ &
\multirow{2}{*}{3.6913}\\
$1^{4}S_{1/2}$ & $\frac{3}{2}^{+}$ & $\Xi_{cc}^{\ast}$ & $n\{cc\}$ & $3.727$ &
\\
$1^{2}S_{1/2}$ & $\frac{1}{2}^{+}$ & $\Xi_{bb}$ & $n\{bb\}$ & $10.202$ &
\multirow{2}{*}{10.2253}\\
$1^{4}S_{1/2}$ & $\frac{3}{2}^{+}$ & $\Xi_{bb}^{\ast}$ & $n\{bb\}$ & $10.237$
& \\
$1^{2}S_{1/2}$ & $\frac{1}{2}^{+}$ & $\Xi_{bc}$ & $n\{bc\}$ & $6.933$ &
\multirow{2}{*}{6.9643}\\
$1^{4}S_{1/2}$ & $\frac{3}{2}^{+}$ & $\Xi_{bc}^{\ast}$ & $n\{bc\}$ & $6.980$ &
\\
$1^{2}S_{1/2}$ & $\frac{1}{2}^{+}$ & $\Xi_{bc}^{\prime}$ & $n[bc]$ & $6.963$ &
6.963\\\hline\hline
\end{tabular}
\end{table}

With the mass parameters $m_{q=n}$ ($n=u, d$) in Table I and the average masses $\bar{M}(1S1s)$ in Table II, one can solve Eq. (\ref{ReggS}) to extract the heavy-diquark mass,
with the results
\begin{align}
M_{cc} &  =2865.5\ \text{MeV, }M_{bb}=8916.7\ \text{MeV, }\label{hdiquark}\\
M_{[bc]} &  =5891.8\ \text{MeV, }M_{\{bc\}}=5892.3\ \text{MeV,}\label{Hdiq}%
\end{align}
where the subscript $cc$($bb$) stands for the vector diquark $\{cc\}$%
($\{bb\}$) for short. The mean mass of the diquark $bc$ is $\bar{M}%
_{bc}=5892.3\ $MeV.

It is examimed in Refs. \cite{JiaLH:19,DPP:Epjc21} for the singly heavy (SH) baryon families(the $\Lambda_{c}/\Sigma_{c}$, the $\Xi_{c}/\Xi_{c}^{\prime}$, the $\Lambda_{b}/\Sigma_{b}$ and the $\Xi_{b}/\Xi_{b}^{\prime}$) that the Regge slope of the baryons are almost same for identical flavor constituents, independent of the light-diquark spin($=0$ or $1$). The trajectory slope is found to rely crucially on the heavy quark mass $M_{Q}$, but has little dependence on the diquark spin of $qq^{\prime}$.

Let us consider first the string tensions of the nonstrange DH baryons $\Xi_{QQ^{\prime}}=nQQ^{\prime}$. Natively, the string tension $a$ are same for all heavy-light systems, namely, the ratio $a_{D}/a_{B}=1=a_{D}/a_{¦®_{QQ}}$ as gluondynamics is flavor-independent, as implied by heavy quark symmetry and heavy quark-diquark (HQD) symmetry at the leading order \cite{HM06}. In the real world, the breaking of the HQD symmetry and heavy quark symmetry yields that the ratios $a_{D}/a_{B}$ and $a_{D}/a_{¦®_{QQ}}$ are not unity but depend upon the respective masses ($M_{c},M_{b}$) and ($M_{c},M_{QQ¡ä}$). The requirement that they tend to unity suggests that this dimensionless ratio depends on some functional of the respective dimensionless ratio $M_{c}/M_{b}$ and $M_{c}/M_{QQ¡ä}$. For simplicity, we use a power-law of the mass scaling between the string tensions of the hadrons $D/B$ and $\Xi_{QQ^{\prime}}$:
\begin{equation}
\frac{a_{D}}{a_{B}}=\left(  \frac{M_{c}}{M_{b}}\right)  ^{P},\label{Rel slope}%
\end{equation}%
\begin{equation}
\frac{a_{D}}{a_{\Xi_{QQ^{\prime}}}}=\left(  \frac{M_{c}}{M_{QQ^{\prime}}%
}\right)  ^{P}.\label{slope}%
\end{equation}
Here, the parameters ($a_{D}=a_{c\bar{n}}$, $a_{B}=a_{b\bar{n}}$, $M_{c}$ and $M_{b}$) are previously evaluated in Ref \cite{JiaLH:19} and listed in Table I. Putting the tensions and heavy-quark masses in Table I into Eq.
(\ref{Rel slope}) gives $P=0.185$ and thereby predicts, by Eq. (\ref{slope})%
\begin{align}
a_{\Xi_{cc}} &  =0.2532\text{ GeV}^{2}\text{, }a_{\Xi_{bb}}=0.3123\text{
GeV}^{2},\label{aXi}\\
a_{\Xi_{bc}} &  =0.2893\text{ GeV}^{2}=a_{\Xi_{bc}^{\prime}}\text{,}%
\label{aXi2}%
\end{align}
combining with Eqs. (\ref{hdiquark}) and (\ref{Hdiq}).

Given the tensions in Eqs. (\ref{aXi}) and Eq. (\ref{aXi2}), and the diquark masses in Eqs. (\ref{hdiquark}), (\ref{Hdiq}) and the light quark mass in Table I, one can use Eq.~(\ref{regge0}) to obtain the spin-averaged masses of the baryon system $\Xi_{QQ^{\prime}}$ in p-wave($l=1$). The results are%
\begin{equation}
\left\{
\begin{array}
[c]{l}%
\bar{M}\left(  \Xi_{cc},1p\right)  =4.0813\mathrm{\ }\text{GeV, }\bar
{M}\left(  \Xi_{bb},1p\right)  =10.5577\mathrm{\ }\text{GeV,}\\
\bar{M}\left(  \Xi_{bc}^{\prime},1p\right)  =7.3258\mathrm{\ }\text{GeV, }%
\bar{M}\left(  \Xi_{bc},1p\right)  =7.3269\mathrm{\ }\text{GeV,}%
\end{array}
\right\}  \label{hdiquark Xi}%
\end{equation}
where $\Xi_{bc}^{\prime}=n[bc]$ and $\Xi_{bc}=n\{bc\}$ stand for the
bottom-charmed baryons with diquark spin $0$ and $1$, respectively.

The same procedure applies to the strange DH baryons, the $\Omega_{QQ^{\prime
}}=sQQ^{\prime}$, and the associated $D_{s}$/$B_{s}$($=c\bar{s}/b\bar{s}$)
mesons, for which the mass scaling, corresponding to Eqs. (\ref{Rel slope})
and (\ref{slope}), has the same form%
\begin{equation}
\frac{a_{D_{s}}}{a_{B_{s}}}=\left(  \frac{M_{c}}{M_{b}}\right)  ^{P}
\label{Rel-slope}%
\end{equation}%
\begin{equation}
\frac{a_{D_{s}}}{a_{\Omega_{QQ^{\prime}}}}=\left(  \frac{M_{c}}{M_{QQ^{\prime
}}}\right)  ^{P}. \label{slop}%
\end{equation}
Putting the tensions for the strange heavy mesons and the heavy-quark masses
in Table I to Eq. (\ref{Rel-slope}) gives $P=0.202$. One can then use Eq. (\ref{slop}) and the heavy-diquark masses in Eqs. (\ref{hdiquark}) as well as (\ref{Hdiq}) to predict%
\begin{align}
a_{\Omega_{cc}}  &  =0.2860\text{ GeV}^{2}\text{, }a_{\Omega_{bb}%
}=0.3596\text{ GeV}^{2}\text{,}\label{aOmi}\\
a_{\Omega_{bc}}  &  =0.3308\ \text{GeV}^{2}=a_{\Omega_{bc}^{\prime}}\text{,}
\label{aOm2}%
\end{align}
where $\Omega_{bc}^{\prime}$ and $\Omega_{bc}$ stand for the DH $\Omega$ baryons
with $bc$-diquark spin $0$ and $1$, respectively.

Given the tensions (including $a_{\Xi_{QQ^{\prime}}}=a_{nQQ^{\prime}}$) in Eqs.
(\ref{aOmi}) and (\ref{aOm2}), and the heavy-diquark masses in Eqs.
(\ref{hdiquark}) and (\ref{Hdiq}) as well as the strange quark mass
$m_{s}=0.328$ GeV in Table I, one can use Eq.~(\ref{regge0}) to find the
mean (spin-averaged) masses of the baryon system $\Omega_{QQ^{\prime}}$ in
p-waves ($l=1$). The results are%
\begin{equation}
\left\{
\begin{array}
[c]{l}%
\bar{M}\left(  \Omega_{cc},1p\right)  =4.1895\mathrm{\ }\text{GeV, }\bar
{M}\left(  \Omega_{bb},1p\right)  =10.6795\mathrm{\ }\text{GeV,}\\
\bar{M}\left(  \Omega_{bc}^{\prime},1p\right)  =7.4430\mathrm{\ }\text{GeV,
}\bar{M}\left(  \Omega_{bc},1p\right)  =7.4441\mathrm{\ }\text{GeV,}%
\end{array}
\right\}  \label{hdiquark Omega}%
\end{equation}
where $\Omega_{bc}^{\prime}=s[bc]$ stands for the strange baryons $\Omega
_{bc}$ with scalar diquark $[bc]$ and $\Omega_{bc}=s\{bc\}$ for the
$\Omega_{bc}$ with axial vector diquark $\{bc\}$.

\renewcommand{\tabcolsep}{0.05cm} \renewcommand{\arraystretch}{1.0}
\begin{table}[h]
\caption{ Spin-averaged(mean) masses of the excited DH baryons
$\Xi_{QQ^{\prime}}$ and $\Omega_{QQ^{\prime}}$ predicted by Eqs. (\ref{ReggeU}),(\ref{dMs}),(\ref{dMd}), with the masses in Eqs. (\ref{hdiquark Xi}),(\ref{hdiquark Omega}). All masses in GeV. }%
\begin{tabular}
[c]{ccccccc}\hline\hline
\textrm{Baryons} & \textrm{$\bar{M}(1S2s)$} & \textrm{$\bar{M}(1S1p)$} &
\textrm{$\bar{M}(2S1s)$} & \textrm{$\bar{M}(1P1s)$} \\\hline
$\Xi_{cc}$ & $4.255$ & $4.081$ & $4.085$ & $4.123$ \\
$\Xi_{bb}$ & $10.722$ & $10.558$ & $10.572$ & $10.630$ \\
$\Xi_{bc}$ & $7.498$ & $7.327$ & $7.330$ & $7.377$ \\
$\Omega_{cc}$ & $4.370$ & $4.190$ & $4.234$ & $4.272$ \\
$\Omega_{bb}$ & $10.854$ & $10.680$ & $10.726$ & $10.783$ \\
$\Omega_{bc}$ & $7.625$ & $7.444$ & $7.482$ & $7.528$ \\\hline\hline
\end{tabular}
\end{table}

To extend the above analysis to radially excited states, one needs a Regge relation to include the radial excitations. For the heavy mesons, this type of Regge relation is proposed in Ref. \cite{JiaD:E19} in consideration
that the trajectory slope ratio between the radial and angular excitations is
$\pi:2$ for the heavy mesons $\bar{q}Q$ in the heavy quark limit. Extending
this ratio to the heavy diquark-quark picture of the DH system $q-QQ^{\prime}$ can be done by viewing the system a QCD string tied to the diquark $QQ^{\prime}$ as one end and to the light quark $q$ at the other. Then, a Regge relation for the radially and angular excitations follows, via utilizing HQD symmetry to
replace heavy quark $Q$ there
by the diquark $\bar{Q}\bar{Q}^{\prime}$ and $\pi
al$ in Eq. (\ref{regge0}) by $\pi a\left(  l+\frac{\pi}{2}n\right)  $, giving
rise to\cite{JiaD:E19}
\begin{align}
\left(  \bar{M}-M_{QQ^{\prime}}\right)  ^{2} &  =\pi a\left(  l+\frac{\pi}%
{2}n\right)  \nonumber\\
&  +\left(  m_{q}+M_{QQ^{\prime}}-\frac{m_{\text{bare }QQ^{\prime}}^{2}%
}{M_{QQ^{\prime}}}\right)  ^{2}.\label{ReggeU}%
\end{align}
This gives a linear Regge relation applicable to the DH anti-baryons $\bar{q}\bar{Q}\bar{Q}^{\prime}$, or the DH baryons $qQQ^{\prime}$.  With the masses of $m_q$ in Table I and the inputs in Eqs. (\ref{mbare}),(\ref{hdiquark}),(\ref{Hdiq}),(\ref{aXi}),(\ref{aXi2}), (\ref{aOmi}) and (\ref{aOm2}), one obtains, by Eq. (\ref{ReggeU}), the mean(spin-averaged) masses of the low-lying DH baryons with 1S2s wave. The results are listed in Table III.

\subsection{The DH baryons with 1P and 2S waves diquark}
Consider excited DH baryons $qQQ^{\prime}$ with diquark $QQ^{\prime}$ in the internal 1P and 2S waves. The diquark $cc$ or $bb$ has to be in spin singlet (asymmetry) in the
internal 1P wave or spin triplet in the internal 2S wave, while $bc$ can be
either in the internal 1P or in 2S waves. As treated before, one can decompose
the excited baryon mass as $M=\bar{M}_{N,L}+(\Delta M)_{N,L}$, where $\bar
{M}_{N,L}$ is the spin-independent mass and $(\Delta M)_{N,L}$ is mass
shift(splitting) due to the spin interactions in DH baryons $qQQ^{\prime}$.

First of all, we consider excited heavy mesons in 1P and 2S waves in QCD
string picture. For an excited heavy quarkonia $Q\bar{Q}$, a rotating-string
picture \cite{BurnsPP:D10} in which a heavy quark at one end and an antiquark at the other with relative orbital momentum $L$, infers a nonlinear Regge relation ($\bar{M}^{3/2}$ $\sim L$), akin
to Eq. (\ref{regge0}),%
\begin{equation}
\bar{M}=2M_{Q}+3\left(  \frac{T^{2}L^{2}}{4M_{Q}}\right)  ^{1/3}. \label{M23}%
\end{equation}

On the other hand, a semiclassical WKB analysis of a system of heavy quarkonia
$Q\bar{Q}$ in a linear confining potential $T|r|$ leads
to a quantization condition for its radial excitations (labeled by $N$,
Appendix B):
\begin{equation}
\left[  \bar{M}-2M_{Q}\right]  ^{3/2}=\frac{3\pi T}{4\sqrt{M_{Q}}}\left(
N+2c_{0}\right)  ,\label{RadN}%
\end{equation}
with the constant $2c_{0}$ the quantum defect. Comparing the radial and
angular slopes (linear coefficients in $N$ and $L$, respectively, which is
$\pi:\sqrt{12}$) of the trajectory in RHS of Eq. (\ref{RadN}) and in that of
Eq. (\ref{M23}), one can combine two trajectories into(Appendix B)
\begin{equation}
\left[  \bar{M}_{N,L}-2M_{Q}-B(Q\bar{Q})_{N,L}\right]  ^{3/2}=\frac{3\sqrt
{3}T}{2\sqrt{M_{Q}}}\left(  L+\frac{\pi N}{\sqrt{12}}+2c_{0}\right)
,\label{MNL}%
\end{equation}
where a term of extra energy $B(Q\bar{Q})_{N,L}$, named heavy-pair binding
energy, enters to represent a corrections to the picture of Regge spectra due
to the short-distance interquark
forces \cite{DeRuju:D75} when two heavy quarks ($Q$ and $\bar{Q}$) come close
each other. Such a term is ignored in the semiclassical picture of massive QCD
string as well as in the WKB analysis (Appendix B). For the ground-state DH
hadrons, a similar binding between two heavy quarks in them was considered in Ref.
\cite{KR:D14,KR:D18}.

For applications in this work, we rewrite Eq. (\ref{MNL}) and extend it to a
general form in which $Q\bar{Q}^{\prime}$ can be $b\bar{c}$ in flavor (by heavy
quark symmetry),
\begin{equation}
\bar{M}(Q\bar{Q}^{\prime})_{N,L}=M_{Q}+M_{Q^{\prime}}+B(Q\bar{Q}^{\prime
})_{N,L}+3\frac{\left[  T\left(  L+\pi N/\sqrt{12}+2c_{0}\right)  \right]
^{2/3}}{[2(M_{Q}+M_{Q^{\prime}})]^{1/3}}.\label{Mbnl}%
\end{equation}
Setting $N=0=L$ in Eq. (\ref{Mbnl}) gives
\begin{equation}
Tc_{0}(Q\bar{Q}^{\prime})=\frac{\sqrt{2(M_{Q}+M_{Q^{\prime}})}}{6\sqrt{3}%
}\left[  \bar{M}(Q\bar{Q})_{1S}-M_{Q}-M_{Q^{\prime}}-B(Q\bar{Q}^{\prime}%
)_{1S}\right]  ^{3/2}.\label{Tc0M}%
\end{equation}
Substitution of the quark masses ($M_{c}=1.44$ GeV, $M_{b}=4.48$ GeV) into Eq. (\ref{Tc0M}) gives (in GeV$^{2}$),
\begin{equation}
Tc_{0}(c\bar{c})=0.0689\text{ GeV}^{2}\text{, } Tc_{0}(b\bar{b})=0.4363\text{
GeV}^{2}\text{,  }Tc_{0}(b\bar{c})=0.2136\text{ GeV}^{2}\text{. }\label{Tc0}%
\end{equation}
in which we have used the observed data in section VI, $\bar{M}(c\bar{c})_{1S}$=3.06865 GeV, $\bar{M}(b\bar{b})_{1S}$=9.4449 GeV, $\bar{M}(b\bar{c})_{1S}$=6.32025 GeV for the ground-state heavy quarkonia and $B_{c}$ meson in Table VIII, and the corresponding binding energies for 1S wave in Table IX,
\begin{equation}
B(c\bar{c})_{1S}=0.25783\text{ GeV}^{2}\text{,} B(b\bar{b})_{1S}=0.56192\text{
GeV}^{2}\text{, } B(b\bar{c})_{1S}=0.3464\text{ GeV}^{2}\text{. }
\end{equation}

Now, we consider the DH baryons $qQQ^{\prime}$ with excited diquark. In the string picture, one can view the baryon as a Y-shaped string system $[Q-Q^{\prime}]-q$ in which a subsystem of massive (rotating or vibrating) string $[Q-Q^{\prime}]$ (tied to one heavy quark $Q$ at each end) is connected via a static string to a light quark $q$ at third end. For 1S or 1P wave diquark in a DH baryon labelled by the quantum numbers $N$ and $L$, respectively, the Regge relation similar to Eq.
(\ref{Mbnl}) takes the form (Appendix C),
\begin{equation}
\bar{M}(QQ^{\prime})_{N,L}=M_{Q}+M_{Q^{\prime}}+\Delta B(QQ^{\prime}%
)_{N,L}+3\frac{\left[  T_{QQ^{\prime}}\left(  (L+\pi N/\sqrt{12}%
)+2c_{0}\right)  \right]  ^{2/3}}{[2(M_{Q}+M_{Q^{\prime}})]^{1/3}}%
+c_{1},\label{MQQb}%
\end{equation}
with $T_{QQ^{\prime}}$ the tension of string of the subsystem $[Q-Q]$ and
$c_{0}$ given by Eq. (\ref{Tc0M}). Here, $c_{1}$ is an additive constant,
defined up to the ground state of the whole DH system.

Since the inverse Regge slope ($1/\alpha^{\prime}$) is derived to scale like $\sqrt{C\alpha_{s}}$ in Ref. \cite{JT}, where $C$ is the Casimir operator and equals to $2/3$ in a color antitriplet $\bar{3}_{c}$ of the pair $QQ^{\prime}$, and $4/3$ in a color singlet ($1_{c}$) of the pair $Q\bar{Q}^{\prime}$, one can write(by $1/\alpha^{\prime}$ $\sim T$)
\begin{equation}
T_{QQ^{\prime}}[\bar{3}_{c}]=\frac{1}{\sqrt{2}}T_{Q\bar{Q}^{\prime}}[1_{c}%
],\label{Thalf}%
\end{equation}
for the heavy pairs with color configuration indicated. So, $\sqrt{2}T_{QQ^{\prime}}c_{0}=T_{Q\bar{Q}^{\prime}}c_{0}$. Accounting the excitation energy of the baryons, $(\Delta\bar{M}^{(\ast
)})_{N,L}=\bar{M}_{N,L}^{(\ast)}-\bar{M}_{0,0}$, for the
energy shift of the subsystem $[Q-Q^{\prime}]$ relative to its ground state, one can write this excitation energy due to diquark excitation ($T_{Q\bar{Q}^{\prime}}=T$ given in Eq. (\ref{Tc0M})) as
\begin{equation}
\bar{M}_{N,L}^{(\ast)}-\bar{M}_{0,0}=\Delta B(QQ^{\prime})_{N,L}%
+3\sqrt{2}\frac{\left[  T(L+\pi N/\sqrt{12})/2+Tc_{0}\right]  ^{2/3}}{[2(M_{Q}%
+M_{Q^{\prime}})]^{1/3}}+C_{1},\label{dMs}%
\end{equation}
where $C_{1}$ is a constant related to $c_{1}$ and determined by(setting
$N=0=L$),
\[
C_{1}[qQQ^{\prime}]=-3\sqrt{2}\left[  \frac{(Tc_{0})^{2}}{2(M_{Q}+M_{Q^{\prime}}%
)}\right]  ^{1/3},
\]
with $\Delta B(QQ^{\prime})_{0,0}=0$. From Eq. (\ref{dMs}) it follows that the baryon mass shift due to diquark excitations becomes
\begin{equation}
(\Delta\bar{M}^{(\ast)})_{N,L}=\Delta B(QQ^{\prime})_{N,L}+3\sqrt{2}\frac{\left[
T(L+\pi N/\sqrt{12})/2+Tc_{0}\right]  ^{2/3}-(Tc_{0})^{2/3}}{[2(M_{Q}%
+M_{Q^{\prime}})]^{1/3}}, \label{dMd}%
\end{equation}
with $Tc_{0}$ given by Eq. (\ref{Tc0}).

Given the parameters in Eqs. (\ref{Tc0}), (\ref{TQQ}) and Tables
\ref{tab:Eff-mass}, II, IX and the mean masses of the ground-state baryons
$\Omega_{cc}$, $\Omega_{bb}$ and $\Omega_{bc}$ in Ref. \cite{EFG:D02}, one can apply Eqs. (\ref{dMs}) and (\ref{dMd}) to find the mean masses of all DH baryons ($\Xi_{cc}$, $\Xi_{bb}$, $\Xi_{bc}$,
$\Omega_{cc}$, $\Omega_{bb}$, $\Omega_{bc}$) with diquark excited radially and angularly. Here, three values ($T_{c\bar{c}}, T_{b\bar{b}}, T_{b\bar{c}}$) of the tension $T$ in Eq. (\ref{dMd}) are obtained in Eq. (\ref{TQQ}) in Section IV via matching Eq. (\ref{Mbnl}) to the mean-mass spectra and binding energies of the heavy quarkonia and $b\bar{c}$ systems. The results for the mean masses of all DH baryons are shown in Table III.

\section{Spin-dependent interactions in $jj$ coupling}
In heavy-diquark quark picture of DH baryon in the ground state (1S1s), two heavy quarks form a S-wave color antitriplet $\left(
\overline{\mathbf{3}}_{c}\right)  $ diquark $(QQ^{\prime})$, having spin
zero ($S_{QQ^{\prime}}=0$) when $QQ^{\prime}=bc$, or spin one ($S_{QQ^{\prime}%
}=1$) when $QQ^{\prime}=cc,bc$ and $bb$. When $Q=Q^{\prime}$ the diquark $QQ$
must have spin one due to full antisymmetry under exchange of two quarks. The
spin $S_{QQ^{\prime}}$ can couple with the spin $S_{q}=1/2$ of the light quark
$q$($=u,d$ and $s$) to form total spin $S_{tot}=1\pm1/2=1/2,3/2$ if
$S_{QQ^{\prime}}=1$ or $S_{tot}=1/2$ if $S_{QQ^{\prime}}=0$.

For the internal ground state of diquark (S-wave), the spin-dependent interaction between light quark and heavy diquark can generally be given by \cite{KR:D15,EFG:D111}
\begin{equation}%
\begin{array}
[c]{c}%
H^{SD}=a_{1}\mathbf{l}\cdot\mathbf{S}_{q}+a_{2}\mathbf{l}\cdot\mathbf{S}%
_{QQ^{\prime}}+bS_{12}+c\mathbf{S}_{q}\cdot\mathbf{S}_{QQ^{\prime}},\\
S_{12}=3\mathbf{S}_{q}\cdot\hat{\mathbf{r}}\mathbf{S}_{QQ^{\prime}}\cdot
\hat{\mathbf{r}}-\mathbf{S}_{q}\cdot\mathbf{S}_{QQ^{\prime}},
\end{array}
\label{spin-d}%
\end{equation}
where the first two terms are spin-orbit forces of the quark $q$ with $\mathbf{S}_{q}$ and the diquark $QQ^{\prime}$ with spin $\mathbf{S}_{QQ^{\prime}}$, the third is a tensor force, and the last describes hyperfine splitting. Here, $l$ stands for the relative orbital angular momentum of $q$ with respective to $QQ^{\prime}$, $\hat{\mathbf{r}} =\vec{\mathbf{r}} / r$ is the unity vector of position pointing from
the center of mass (CM) of the diquark to the light quark q.

For the 1Sns waves of the DH baryons ($\mathbf{l}=0$), only the last term survives in Eq. (\ref{spin-d}), in which $\mathbf{S}_{QQ^{\prime}}\cdot\mathbf{S}_{q}$ has the eigenvalues $\{-1,1/2\}$ when $\mathbf{S}_{QQ^{\prime}}=1$. The mass splitting
for the systems $q\{QQ^{\prime}\}$ becomes ($J=1/2,3/2$),
\begin{equation}
\Delta M(q\{QQ^{\prime}\})=c(q\{QQ^{\prime
}\}) \left[
\begin{array}
[c]{rr}%
-1 & 0\\
0 & 1/2
\end{array}
\right]  \text{.}\label{2Sm}
\end{equation}

For the 1Snp waves of the bottom charmed baryons $\Xi_{bc}$ and $\Omega_{bc}$ with zero diquark
spin $S_{bc}=0$, the spin interaction (\ref{spin-d}) yields the mass splitting ($J=1/2,3/2$),
\begin{equation}
\Delta M(q[QQ^{\prime}])=a_{1}(q[QQ^{\prime}])\left[
\begin{array}
[c]{rr}%
-1 & 0\\
0 & 1/2
\end{array}
\right]  \text{.} \label{Mbc}%
\end{equation}

For the case of excited(2S or 1P wave) diquark $QQ^{\prime}$, a correction to Eq. (\ref{spin-d}) emerges due to the interaction between the diquark and light quark, given by
\begin{equation}
H^{dSD}=c^{\ast}(\mathbf{L+S}_{QQ^{\prime}})\cdot\mathbf{S}_{q}\text{,}%
\label{HLss}%
\end{equation}
where the diquark spin $\mathbf{S}_{QQ^{\prime}}=1$ (S wave of diquark) or $0$ (P wave) when $QQ^{\prime}=cc$ or $bb$. This correction stems from the interaction of the effective magnetic moments $e_{QQ^{\prime}}(\mathbf{L+S}_{QQ^{\prime}})/(2M_{QQ^{\prime}})$ of the excited diquark and the spin magnetic moment $e_{q}\mathbf{S}_{q}/m_{q}$ of the light quark $q$. Here, $\mathbf{L}$ is the internal orbital angular
moment of the diquark, $e_{QQ^{\prime}}$ and $e_{q}$ stand for the respective charges of diquark and light quark.

For the $bc$-diquark for which $QQ^{\prime}=[bc]$ or $\{bc\}$, the spin $\mathbf{S}_{[bc]}=0$ in internal $S$ wave and $\mathbf{S}_{[bc]}=1$ in internal P wave by the symmetry of the baryon states
$q[bc]\ $and $q\{bc\}$. Also, $\mathbf{S}_{\{bc\}}=1$ in internal $S$ wave and
$\mathbf{S}_{\{bc\}}=0$ in internal P wave. For the baryon $q[bc]$, $\mathbf{J}%
_{d}=\mathbf{L\oplus S}_{QQ^{\prime}}$ $=\{0\}$ in S wave of diquark or
$\{0,1,2\}$ in P wave of diquark, while for the baryon $q\{bc\}$,
$\mathbf{J}_{d}=\mathbf{L\oplus S}_{QQ^{\prime}}$ $=\{1\}$ in the 2S and 1P
wave of the diquark.

\subsection{The 1p wave DH baryons with S wave diquark}

Consider now the excited (1p) DH baryon in which light quark is excited to 1p wave ($l=1$) relative to $QQ^{\prime}$. Note that coupling $S_{tot}=1/2$ to $l=1$ gives the states with the total angular momentums $J=1/2,3/2$, while coupling $S_{tot}=3/2$ to $l=1$ leads to the states with $J=1/2,3/2$ and $5/2$. We use then the $LS$ basis $^{2S_{tot}+1}P_{J}$ $=\{^{2}P_{1/2},^{2}P_{3/2},^{4}P_{1/2},^{4}P_{3/2},^{4}P_{5/2}\}$ to label these
multiplets in p-wave ($J=1/2,1/2^{\prime},3/2,3/2^{\prime}$ and $5/2$). The two $J=1/2$ states are the respective eigenstates of a $2\times2$ matrices $\mathcal{M}_{J}$ representing $H^{SD}$ for $J=1/2$ and $3/2$. In terms of the basis $[^{2}P_{J},^{4}P_{J}]$, they can be given by the matrix \cite{KR:D2017,Landau}(see Appendix D also)%
\begin{align}
\mathcal{M}_{J=1/2}  &  =\left[
\begin{array}
[c]{ll}%
\frac{1}{3}\left(  a_{1}-4a_{2}\right)  & \frac{\sqrt{2}}{3}\left(
a_{1}-a_{2}\right) \\
\frac{\sqrt{2}}{3}\left(  a_{1}-a_{2}\right)  & -\frac{5}{3}\left(
a_{2}+\frac{1}{2}a_{1}\right)
\end{array}
\right] \nonumber\\
&  +b\left[
\begin{array}
[c]{cc}%
0 & \frac{\sqrt{2}}{2}\\
\frac{\sqrt{2}}{2} & -1
\end{array}
\right]  +c\left[
\begin{array}
[c]{cc}%
-1 & 0\\
0 & \frac{1}{2}%
\end{array}
\right]  , \label{J1/2}%
\end{align}
in the $J=1/2$ subspace,
\begin{align}
\mathcal{M}_{J=3/2}  &  =\left[
\begin{array}
[c]{cc}%
-\frac{2}{3}\left(  \frac{1}{4}a_{1}-a_{2}\right)  & \frac{\sqrt{5}}{3}\left(
a_{1}-a_{2}\right) \\
\frac{\sqrt{5}}{3}\left(  a_{1}-a_{2}\right)  & -\frac{2}{3}\left(  \frac
{1}{2}a_{1}+a_{2}\right)
\end{array}
\right] \nonumber\\
&  +b\left[
\begin{array}
[c]{cc}%
0 & -\frac{\sqrt{5}}{10}\\
-\frac{\sqrt{5}}{10} & \frac{4}{5}%
\end{array}
\right]  +c\left[
\begin{array}
[c]{cc}%
-1 & 0\\
0 & \frac{1}{2}%
\end{array}
\right]  , \label{J3/2}%
\end{align}
in the $J=3/2$ subspace, and by
\begin{equation}
\mathcal{M}_{J=5/2}=\frac{1}{2}a_{1}+a_{2}-\frac{b}{5}+\frac{c}{2}.
\label{J5/2}%
\end{equation}
for the $J=5/2$. One can verify that the spin-weighted sum of these matrixes
over $J=1/2,3/2$ and $5/2$ is zero:
\begin{equation}
\sum_{J}(2J+1)\mathcal{M}_{J}=0, \label{sumJ}%
\end{equation}
as it should be for the spin-dependent interaction $H^{SD}$.

In the heavy quark limit ($M_{Q}\rightarrow\infty$), all terms except for the first ($=$ $a_{1}\mathbf{l}\cdot\mathbf{S}_{q}$) in
Eq. (\ref{spin-d}) behave as $1/M_{QQ^{\prime}}$ and are suppressed. Due to heavy quark spin symmetry ($\mathbf{S}_{QQ^{\prime}}$ conserved), the total angular momentum of the light quark $\mathbf{j}=\mathbf{l}+\mathbf{S}_{q}\mathbf{=J-S}%
_{QQ^{\prime}}$ is conserved and forms a set of the conserved operators
\{$\mathbf{J},\mathbf{j}$\}, where $\mathbf{J}$ is the total angular momentum
of the DH hadrons. We use then the $jj$ coupling scheme to label
the spin multiplets of the DH baryons, denoted by the basis $|J,$ $j\rangle
$(Appendix D), in which the spin of diquark $QQ^{\prime}$ decouples and
$\mathbf{l}\cdot\mathbf{S}_{q}$ becomes diagonal. As such, the formula for
mass splitting $\Delta M$ can be obtained by diagonalizing $ a_{1}\mathbf{l}%
\cdot\mathbf{S}_{q}$ and treating other interactions in Eq. (\ref{spin-d}) perturbatively.

The eigenvalues (two diagonal elements) of $\mathbf{l}\cdot\mathbf{S}_{q}$ can
be obtained to be
\begin{align}
\left\langle \mathbf{l}\cdot\mathbf{S}_{q}\right\rangle  &  =\frac{1}%
{2}\left[  j(j+1)-l(l+1)-S_{q}\left(  S_{q}+1\right)  \right]  \nonumber\\
&  =-1(j=1/2)\text{ or }1/2(j=3/2).\label{LSq}%
\end{align}
Thus, one can compute mass splitting $\Delta M(J,j)=\left\langle
J,j\left\vert H^{SD}\right\vert J,j\right\rangle $ in the $jj$ coupling
via three steps: Firstly, one solves the eigenfunctions(the $LS$
bases) $\left\vert S_{QQ^{\prime}3},S_{q3},l_{3}\right\rangle $ of the
$\mathbf{l}\cdot\mathbf{S}_{q}$ for its eigenvalues $j=1/2$ and $3/2$ in the
$LS$ coupling; Secondly, one transforms the obtained $LS$ bases into
$\left\vert J,j\right\rangle $ in the $jj$ coupling (Appendix D); Finally, one
evaluates the diagonal elements of the spin-interaction in Eq.~(\ref{spin-d})
in the new bases $\left\vert J,j\right\rangle $. The results for $\Delta
M(J,j)$ are listed in Table IV. \renewcommand{\tabcolsep}{0.45cm}
\renewcommand{\arraystretch}{1.0}
\begin{table}[tbh]
\caption{The matrix elements of the mass splitting operators in the p-wave doubly heavy baryons states in the $jj$ coupling.}%
\label{tab:matrix elements}
\begin{tabular}
[c]{cccc}\hline\hline
\textrm{(J-j)} & \textrm{$\left\langle \mathbf{l}\cdot\mathbf{S}_{QQ^{\prime}%
}\right\rangle $} & \textrm{$\left\langle \mathbf{S}_{12}\right\rangle $} &
\textrm{$\left\langle \mathbf{S}_{q}\cdot\mathbf{S}_{QQ^{\prime}}\right\rangle
$}\\\hline
$(1/2,1/2)$ & $-4/3$ & $-4/3$ & $1/3$\\
$(1/2,3/2)$ & $-5/3$ & $1/3$ & $-5/6$\\
$(3/2,1/2)$ & $2/3$ & $2/3$ & $-5/6$\\
$(3/2,3/2)$ & $-2/3$ & $2/15$ & $-1/3$\\
$(5/2,3/2)$ & $1$ & $-1/5$ & $1/2$\\\hline\hline
\end{tabular}
\end{table}

Given Table III, one can use the lowest perturbation theory to find
\begin{equation}
\Delta M(1/2,1/2)=-a_{1}-\frac{4}{3}a_{2}-\frac{4}{3}b+\frac{1}{3}%
c,\label{M11}%
\end{equation}%
\begin{equation}
\Delta M(1/2,3/2)=\frac{1}{2}a_{1}-\frac{5}{3}a_{2}+\frac{1}{3}b-\frac{5}%
{6}c,\label{M13}%
\end{equation}%
\begin{equation}
\Delta M(3/2,1/2)=-a_{1}+\frac{2}{3}a_{2}+\frac{2}{3}b-\frac{1}{6}%
c,\label{M31}%
\end{equation}%
\begin{equation}
\Delta M(3/2,3/2)=\frac{1}{2}a_{1}-\frac{2}{3}a_{2}+\frac{2}{15}b-\frac{1}%
{3}c,\label{M33}%
\end{equation}%
\begin{equation}
\Delta M(5/2,3/2)=\frac{1}{2}a_{1}+a_{2}-\frac{1}{5}b+\frac{1}{2}c,\label{M53}%
\end{equation}
which express the baryon mass splitting in p wave in terms of four parameters
($a_{1},a_{2},b,c$). The mass formula for the $1S1p$
states is then $M(J,j)=\bar{M}(1S1p)+\Delta
M(J,j)$, with $\bar{M}(1S1p)$ the spin-independent masses given in Eqs.
(\ref{hdiquark Xi}) and (\ref{hdiquark Omega}) in section II.

\subsection{The DH baryons with 2S and 1P wave diquark}
In the heavy diquark-quark picture in this work, diquark $QQ$ is fundamentally two body system connected by string and can be excited internally. In this subsection, we consider the spin interaction due to the 2S and 1P wave excitations of diquark in DH baryons.

(1) The $2S1s$ states. In this state, $\mathbf{L}=0$ and the spin-interaction (\ref{HLss}) reduces to $H^{SD}(2S)=c^{\ast}\mathbf{S}_{QQ^{\prime}}\cdot\mathbf{S}_{q}$, in which $\mathbf{S}_{QQ^{\prime}}\cdot\mathbf{S}_{q}$ has the eigenvalues $\{-1,1/2\}$ for the spin of total system $J=1/2$ or $3/2$, respectively. Note that we occasionally use $2S$ to stand for $2S1s$ for short. Here, the excited baryon energy stems from the spin interaction $H^{SD}(2S)$ as well as the string energy shift $(\Delta\bar{M}^{(\ast)})_{2S}$ given in Eq. (\ref{dMd}). Thus, the mass splitting for the $2S1s$ wave baryons $q\{QQ^{\prime}\}$ becomes($J=1/2$, $3/2$),
\begin{equation}
\Delta M(q\{QQ^{\prime}\})_{2S}=(\Delta\bar{M}^{(\ast)})_{2S}+c^{\ast
}(q\{QQ^{\prime}\})\left[
\begin{array}
[c]{rr}%
-1 & 0\\
0 & 1/2
\end{array}
\right]  \text{.}\label{SSM}
\end{equation}

In the case of $2S1s$ wave DH system $q[bc]$, no mass splitting happens since $\mathbf{L}=0=\mathbf{S}_{[bc]}$, namely, the baryon mass is simply $\bar{M}(q[bc])_{nS}$.

(2) The $1P1s$ states. For the systems $q\{cc\}$, $q\{bb\}$ and $q\{bc\}$,
$\mathbf{S}_{QQ^{\prime}}=0$ and the system spin $\mathbf{J}=\mathbf{L}%
\oplus1/2$ takes values $J=1/2$ and $3/2$. The spin interaction becomes
$c^{\ast}\mathbf{L}\cdot\mathbf{S}_{q}$, which equals to $c^{\ast
}diag[-1,1/2]$. So, the baryon mass splitting is($J=1/2$, $3/2$),
\begin{equation}
\Delta M(q\{QQ^{\prime}\})_{1P}=(\Delta\bar{M}^{(\ast)})_{1P}+c^{\ast
}\left[
\begin{array}
[c]{rr}%
-1 & 0\\
0 & 1/2
\end{array}
\right]  \text{.}\label{SSM1}
\end{equation}

For the systems $q[bc]$, $\mathbf{S}_{QQ^{\prime}}=1$, and $\mathbf{J}%
_{d}=\mathbf{L\oplus S}_{QQ^{\prime}}$ can take values $J_{d}=0$, $1$ or $2$
so that $J=1/2,1/2^{\prime},3/2,3/2^{\prime},5/2$. Labelling the DH baryon systems by $|J,J_{d}\rangle$, the relation $\mathbf{J}_{d}\cdot
\mathbf{S}_{q}=[J(J+1)-J_{d}(J_{d}+1)-3/4]/2$ yields the mass splitting matrix for the P wave multiplets,
\begin{equation}
\Delta M(q[bc])_{1P}=(\Delta\bar{M}^{(\ast)})_{1P}+c^{\ast}diag\left[
0,-1,\frac{1}{2},-\frac{3}{2},1\right]  ,\label{LssM}%
\end{equation}
in the subspace of \{$|J,J_{d}\rangle$\}$=$\{$|1/2,0\rangle$, $|1/2,1\rangle$,
$|3/2,1\rangle$,$|3/2,2\rangle$, $|5/2,2\rangle$\}.

\section{Spin couplings of the DH baryons via mass scaling}

To evaluate spin coupling parameters in Eq.~(\ref{M11}) through Eq.~(\ref{LssM}) for the excited DH baryons, we utilize, in this section, the relations of mass
scaling, which apply successfully to the SH hadrons\cite{KR:D15,JiaLH:19}. For this, we list, in Table V, the experimentally matched values of the spin couplings (data before parentheses) for the existing heavy-light systems, such as the $D_{s}$ in Refs. \cite{KR:D15,JiaLH:19} and the SH baryons
(the $\Sigma_{Q}$/$\Xi_{Q}^{\prime}$/$\Omega_{Q}$, with $Q=c,b$)
in Ref. \cite{DPP:Epjc21}.

\renewcommand{\tabcolsep}{0.22cm} \renewcommand{\arraystretch}{1.0}
\begin{table}[tbh]
\caption{Effective masses (in GeV) of light diquarks in the singly heavy
baryons shown. Data are from the Ref. \cite{JiaLH:19} and Ref.
\cite{DPP:Epjc21}. }%
\begin{tabular}
[c]{ccccccc}\hline\hline
\textrm{Hadrons} & \textrm{$\Sigma_{c}$} & \textrm{$\Sigma_{b}$} &
\textrm{$\Xi_{c}^{\prime}$} & \textrm{$\Xi_{b}^{\prime}$} & \textrm{$\Omega
_{c}$} & \textrm{$\Omega_{b}$}\\\hline
$m_{qq}$ & $0.745$ & $0.745$ & $0.872$ & $0.872$ & $0.991$ & $0.991$%
\\\hline\hline
\end{tabular}
\end{table}

\subsection{The 1p wave baryons with 1S wave diquark }

Before considering DH baryons, let us first examine the mass scaling of the spin couplings between heavy baryons $Qqq^{\prime}$ and heavy mesons(e.g., the
$D_{s}=c\bar{s}$). In Refs. \cite{KR:D15,JiaLH:19}, a relation of mass scaling is explored based on Breit-Fermi like interaction \cite{MNS,Bali01}, and is given by
\cite{JiaLH:19}(see Eqs. (22-24))
\begin{equation}
\mathbf{l}_{qq^{\prime}}\cdot\mathbf{S}_{qq^{\prime}}:a_{1}\left[
Qqq^{\prime}\right]  =\frac{M_{c}}{M_{Q}}\cdot\frac{m_{s}}{m_{qq^{\prime}}%
}\cdot a_{1}\left(  D_{s}\right)  ,\label{a1}%
\end{equation}%
\begin{equation}
\mathbf{l}_{qq^{\prime}}\cdot\mathbf{S}_{Q}:a_{2}\left[  Qqq^{\prime}\right]
=\frac{M_{c}}{M_{Q}}\cdot\frac{1}{1+m_{qq^{\prime}}/M_{c}}\cdot a_{2}\left(
D_{s}\right)  ,\label{a2}%
\end{equation}%
\begin{equation}
S_{12}:b\left[  Qqq^{\prime}\right]  =\frac{M_{c}}{M_{Q}}\cdot\frac
{1}{1+m_{qq^{\prime}}/m_{s}}\cdot b\left(  D_{s}\right)  ,\label{b}%
\end{equation}
where $\mathbf{l}_{qq^{\prime}}$ and $\mathbf{S}_{qq^{\prime}}$ denote the
orbital angular momentum of light diquark $qq^{\prime}$ relative to the heavy
quark $Q$ and the spin of the light diquark, respectively, $m_{qq^{\prime}}$ is the diquark mass and $M_{Q}$ the heavy-quark mass in the SH hadrons. The factor $M_{c}/M_{Q}$ enters to account for the heavy quark dependence. In Eq. (\ref{b}) , the extra recoil factor $1/\left(  1+m_{qq^{\prime}}/m_{s}\right) $ enters to take into account the correction due to comparable heaviness between the diquark $qq^{\prime}$ and the strange quark.
Note that a similar (recoil) factor, $1/\left(  1+m_{qq^{\prime}}%
/M_{c}\right)  $, entering Eq. (\ref{a2}), has been confirmed for the charmed
and bottom baryons in P-wave and D-wave\cite{JiaLH:19}. For instance, Eq.
(\ref{a2}) in Ref. \cite{JiaLH:19}(Eq. (60)) well reproduces the measured
masses $6146.2$ MeV and $6152.5$ MeV of the $\Lambda_{b}(6146)$ and the
$\Lambda_{b}(6152)$ observed by LHCb\cite{Lamdab19}. Similar
verifications were demonstrated in Ref. \cite{DPP:Epjc21} for the excited
$\Omega_{c}$ discovered by LHCb\cite{Aaij:L2017}, for which the $ss$-diquark
is comparable with the charm quark in heaviness.

\renewcommand{\tabcolsep}{0.30cm}
\renewcommand{\arraystretch}{1.0}
\begin{table}[tbh]
\caption{Spin-coupling parameters(in MeV) of the heavy meson $D_{s}$
and the SH baryons $\Sigma_{Q}$, $\Xi_{Q}^{\prime}$ and $\Omega_{Q}$($Q=c,b$). The data before parentheses are the parameters matched to the measured spectra of the SH baryons shown and the data in parentheses are that computed by the mass scaling Eqs. (\ref{a1}), (\ref{a2}) and (\ref{b}). }
\label{tab:sc parameters}
\begin{tabular}
[c]{cccc}\hline\hline
\textrm{Hadrons} & \textrm{$a_{1}$} & \textrm{$a_{2}$} & \textrm{$b$}\\\hline
$D_{s}$ & $89.36$ & $40.7$ & $65.6$\\
$\Sigma_{c}$ & $39.96(39.34)$ & $21.75(26.82)$ & $20.70(20.05)$\\
$\Sigma_{b}$ & $12.99(12.65)$ & $6.42(8.62)$ & $6.45(6.45)$\\
$\Xi_{c}^{\prime}$ & $32.89(33.62)$ & $20.16(25.35)$ & $16.50(17.93)$\\
$\Xi_{b}^{\prime}$ & $9.37(10.81)$ & $6.29(8.15)$ & $5.76(5.76)$\\
$\Omega_{c}$ & $26.96(29.59)$ & $25.76(24.11)$ & $13.51(16.31)$\\
$\Omega_{b}$ & $8.98(9.51)$ & $4.11(7.75)$ & $7.61(5.24)$\\\hline\hline
\end{tabular}
\end{table}

Putting the parameters in Tables I and the light diquarks in
Table V to Eqs.~(\ref{a1})-(\ref{b}), one can estimate the parameters
$a_{1},a_{2}$ and $b$ for the heavy baryons $\Sigma_{Q}$, the $\Xi_{Q}^{^{\prime}}$ and the $\Omega_{c}$. In Table VI, we list the obtained results within parentheses so that they are comparable with that (data before parentheses) matched to the measured data in Ref. \cite{DPP:Epjc21}.
Evidently, the mismatch shown is small: $\Delta a_{1}\leq2.63$ MeV, $\Delta
a_{2}\leq5.19$ MeV and $\Delta b\leq2.80$ MeV, and agreement is noticeable .

Consider the DH baryons with S-wave diquark now. Regarding mass scaling between the DH baryons($QQq$) and the heavy mesons($Q\bar{q}$), the heavy diquark-antiquark(HDA) duality or symmetry($QQ
 \leftrightarrow \bar Q$) in Refs. \cite{SW90,HM06,BVK05} suggests that two hadrons share the same chromodynamics in the heavy quark limit($M_{Q}\rightarrow \infty $) up to a color factor. In the real world where $M_{Q}$ is finite, this symmetry(duality) breaks and the dynamics degenerates to a similar dynamics which mainly depends on the similarity(asymmetric) parameter $M_{Q}/M_{QQ}$. As two heavy quarks in diquark $QQ^{\prime}$ moves in relatively smaller region ($\propto 1/M_{Q}$ ), compared to the DH baryon itsef($\propto 1/m_{q}$ ), the string structure of
the $q-QQ^{\prime}$ resembles that of the heavy meson $D_{s}$, one naturally expects that similar relations of mass scaling apply to the coupling parameters in Eq. (\ref{spin-d}) for the DH baryons.

Replacing $M_{Q}$ in Eqs.~(\ref{a1})-(\ref{b}) by the diquark mass $M_{QQ^{\prime}}$, and the  mass of the light diquark mass $m_{qq^{\prime}}$ by
that of the light quark, one obtains
\begin{equation}
a_{1}[QQ^{\prime}q]=\frac{M_{c}}{M_{QQ^{\prime}}}\cdot\frac{m_{s}}{m_{q}}\cdot
a_{1}\left(  D_{s}\right)  , \label{A11}%
\end{equation}%
\begin{equation}
a_{2}[QQ^{\prime}q]=\frac{M_{c}}{M_{QQ^{\prime}}}\cdot\frac{1}{1+m_{q}/M_{c}%
}\cdot a_{2}\left(  D_{s}\right)  , \label{A22}%
\end{equation}%
\begin{equation}
b[QQ^{\prime}q]=\frac{M_{c}}{M_{QQ^{\prime}}}\cdot\frac{1}{1+m_{q}/m_{s}}\cdot
b\left(  D_{s}\right)  . \label{B}%
\end{equation}

Since the hyperfine parameter $c\propto |\psi_{B}(0)|^{2}/(M_{QQ^{\prime}}m_{q})$ , scales as hadron wavefunction $|\psi_{B}(0)|^{2}$ near the origin, it should be small in p-wave. One can write a relation of mass scaling relative to the singly charmed baryon $\Omega_{c}=css$ as below:
\begin{equation}
c[QQ^{\prime}q]=\left(  \frac{M_{c}}{M_{QQ^{\prime}}}\right)  \left(
\frac{m_{ss}}{m_{q}}\right)  c\left(  \Omega_{c}\right)  . \label{c}%
\end{equation}
Experimentally, there exists five excited $\Omega_{c}$'s discovered by
LHCb\cite{Aaij:L2017}, with the masses $3000.4$ MeV, $3050.2$ MeV, $3065.6$
MeV, $3095.2$ MeV and $3119.2$ MeV. Assigning the five states to be in
p-wave and matching the $jj$ mass formula to the five measured masses lead to
\cite{KR:D2017,DPP:Epjc21}
\begin{equation}
c\left(\Omega_{c}\right)  =4.07\text{ MeV, } m_{ss}=991\text{ MeV. } \label{cOm}%
\end{equation}

Now, one can employ Eqs.~(\ref{A11})-(\ref{c}) to estimate the four parameters
of spin couplings of the DH baryons using the parameters in Table I and the diquark masses in Eqs. (\ref{hdiquark}), (\ref{Hdiq}) and that
in Eq. (\ref{cOm}). The results are listed collectively in Table VII.
\renewcommand{\tabcolsep}{0.42cm} \renewcommand{\arraystretch}{1.0}
\begin{table}[tbh]
\caption{Spin-coupling parameters(in MeV) in the spin interaction
(\ref{spin-d}) of the DH baryons $\Xi_{QQ^{\prime}}$ and $\Omega_{QQ^{\prime}%
}$($Q,Q^{\prime}=c,b$) in p wave.}%
\label{tab:dh baryons}
\begin{tabular}
[c]{ccccc}\hline\hline
DH baryons & \textrm{$a_{1}$} & \textrm{$a_{2}$} & \textrm{$b$} & \textrm{$c$%
}\\\hline
$\Xi_{cc}$ & $64.05$ & $17.64$ & $19.38$ & $8.81$\\
$\Xi_{bb}$ & $20.58$ & $5.67$ & $6.23$ & $2.83$\\
$\Xi_{\{bc\}}$ & $31.14$ & $8.58$ & $9.42$ & $4.29$\\
$\Omega_{cc}$ & $44.91$ & $16.66$ & $16.48$ & $6.18$\\
$\Omega_{bb}$ & $14.43$ & $5.35$ & $5.30$ & $1.99$\\
$\Omega_{\{bc\}}$ & $21.84$ & $8.10$ & $8.01$ & $3.01$\\\hline\hline
\end{tabular}
\end{table}
As seen in Table, the parameter $a_{2}$ are smaller(three times
roughly) but comparable to $a_{1}$. The magnitudes of $b$ is as large as
$a_{2}$ roughly while $c$ is the smallest. This agrees qualitatively with the
parameter hierarchy of the spin-couplings\cite{DPP:Epjc21,JiaLH:19} implied by
heavy quark symmetry.

With $a_{1}(D_{s})=89.4$ MeV in Table VI\cite{KR:D15}, one can employ the relations of
mass scaling (\ref{A11}) relative to the $D_{s}$ to give
\begin{align}
a_{1}(\Xi_{bc})  &  =\left(  \frac{1.440\text{ GeV}}{5.8918\text{ GeV}%
}\right)  \left(  \frac{0.328\text{ GeV}}{0.230\text{ GeV}}\right)
a_{1}(D_{s})=31.14\text{ MeV,}\nonumber\\
a_{1}(\Omega_{bc})  &  =\left(  \frac{1.440\text{ GeV}}{5.8918\text{ GeV}%
}\right)  a_{1}(D_{s})=21.84\text{ MeV,} \label{a1DH}%
\end{align}
where $M_{[bc]}=5.8918$ GeV and other mass parameters are from Table I.
Putting above parameters into Eq. (\ref{Mbc}), and adding $\bar{M}_{L}%
(\Xi_{bc}$,$\Omega_{bc})$ in Eqs. (\ref{hdiquark Xi}) and
(\ref{hdiquark Omega}), one obtains the P-wave masses $\bar{M}_{L}(\Xi_{bc}%
$,$\Omega_{bc})+\Delta M(\Xi_{bc}$,$\Omega_{bc})$ for the baryons $\Xi_{bc}$
and $\Omega_{bc}$(with scalar diquark $bc$ and $J^{P}=1/2^{-}$, $3/2^{-}%
$) to be%
\begin{align}
M\left(  \Xi_{bc}\text{,}1/2^{-}\right)   &  =7294.7\text{ MeV, }M\left(
\Xi_{bc}\text{,}3/2^{-}\right)  =7341.4\text{ MeV,}\label{Xibc}\\
M\left(  \Omega_{bc}\text{,}1/2^{-}\right)   &  =7421.2\text{ MeV, }M\left(
\Omega_{bc}\text{,}3/2^{-}\right)  =7453.9\text{ MeV. } \label{Ombc}%
\end{align}

We list all masses of the DH baryons in their 1S1p-wave in Tables X to XVI,
with comparison with other predictions cited. These masses are also shown in FIGs 1-8 for the respective states of the DH baryons. Our computation suggests that the 1S1p states of the DH baryons have increasing masses with baryon spin $J$ rising from the lowest $1/2^{-}$ to highest $5/2^{-}$.

\subsection{The 2s and 1s wave baryons with 1S wave diquark}
We consider first the DH baryons $qQQ^{\prime}$ with excited diquark $QQ^{\prime}$ in internal 2S waves and then examine the DH baryons in ground state. As two spin-states of the $D$ mesons are established in 1s and 2s waves experimentally, we shall use the mass scaling relative to the $D$ mesons.

In 2s wave($n=1$), two $D$ mesons are the $D_{0}(2550)^{0}$ with mass $2549\pm19$ MeV and the $D_{1}^{\ast}(2600)^{\pm0}$ with mass $2627\pm10$ MeV\cite{ZylaPDG:D20}). The mass scaling relative to the $D$ mesons, inspired by Breit-Fermi like interaction \cite{MNS,Bali01}, can be given by
\begin{equation}
c\left(QQ^{\prime}u\right)_{2s}=\left(  \frac{M_{c}}{M_{QQ^{\prime}}}\right)  c\left( D\right)_{2s} . \label{c2s}%
\end{equation}
It follows ( $c(D)_{2s}=$ mass spliting of the heavy mesons) that
\begin{align*}
c\left(  \Xi_{bb}(bbu)\right)   &  =\left(  \frac{M_{c}}{M_{bb}}\right)
\left[  M(D(2s),1^{-}))-M\left(  D(2s),0^{-}\right)  \right]  =12.6\text{
MeV,}\\
c\left(  \Omega_{bb}(bbs)\right)   &  =\left(  \frac{M_{c}}{M_{bb}}\right)
\left(  \frac{m_{u}}{m_{s}}\right)  \left[  M(D(2s),1^{-})-M\left(
D(2S),0^{-}\right)  \right]  =8.8\text{ MeV.}%
\end{align*}

Further application of mass scaling between the DH baryons,
\begin{align}
c\left(  \Xi_{cc}\right)  _{2s}  &  =c\left(  \Xi_{bb}\right)  \left(
\frac{M_{bb}}{M_{cc}}\right)  ,c\left(  \Xi_{bc}\right)  _{2s}=c\left(
\Xi_{bb}\right)  \left(  \frac{M_{bb}}{M_{bc}}\right)  ,\nonumber\\
c\left(  \Omega_{cc}\right)_{2s}  &  =c\left(  \Omega_{bb}\right)  \left(
\frac{M_{bb}}{M_{cc}}\right), c\left(  \Omega_{bc}\right)_{2s}  =c\left(
\Omega_{bb}\right)  \left(  \frac{M_{bb}}{M_{bc}}\right), \label{coc}%
\end{align}
leads to the following parameter $c$ for the 2s-wave baryons $\Xi_{cc}$,
$\Xi_{bc}$, $\Omega_{cc}$ and $\Omega_{cc}$,$\allowbreak$
\begin{align}
c(\Xi_{cc})_{2s}  &  =39.2\text{ MeV, }c\left(  \Xi_{bc}\right)
_{2S}=19.1\text{ MeV,}  \nonumber\\
c(\Omega_{cc})_{2s}&=27.4\text{ MeV, } c\left(\Omega_{bc}\right)_{2s}  =13.3\text{ MeV.} \label{ccc}%
\end{align}

With the mean(2s) masses in Table III, one can employ Eq. (\ref{2Sm}) to compute the 2s-wave(namely, the $1S2s$-wave) masses of the DH baryons. The results are listed in Table X through Table XVI, with comparison with other calculations, and alson shown in FIGs 1-8, respectively.

It is of interest to apply the same strategy to the ground states to check if
one can reproduce the masses in Table II. For the 1s wave $\Xi_{cc}$ firstly,
the $c$ value can be scaled to that of the ground state $D$ mesons with mass
difference $m(D^{\ast}(2010)^{+})-m(D^{+})=140.6$ MeV\cite{ZylaPDG:D20} via
mass scaling:
\[
c\left(  \Xi_{cc}\right)  _{1s}=c\left(  D\right)_{1s}\left(  \frac{M_{c}%
}{M_{cc}}\right)  =(140.6\ \text{MeV})\left(  \frac{1.44}{2.8655}\right)
=70.7\ \text{MeV,}%
\]
For the $\Omega_{cc}$, the relevant meson for scaling is the ground-state
$D_{s}$ meson with mass difference $m(D_{s}^{\ast\pm})-m(D_{s}^{\pm})=143.8$
MeV\cite{ZylaPDG:D20}, and the $c$ value is then
\[
c\left(  \Omega_{cc}\right)  _{1s}=c\left(  D_{s}\right)  _{1s}\left(
\frac{M_{c}}{M_{cc}}\right)  =(143.8\ \text{MeV})\left(  \frac{1.44}%
{2.8655}\right)  =72.3\ \text{MeV.}%
\]

Using $\bar{M}(qQQ^{\prime})_{1s}=M_{QQ^{\prime}}+m_{q}+M_{QQ^{\prime}%
}v_{QQ^{\prime}}^{2}$ given by Eq. (\ref{ReggS}) with $v_{QQ^{\prime}}^{2}\equiv1-$$m_{\text{bare}QQ^{\prime}}^{2}/M_{QQ^{\prime}}^{2}$ and Eq. (\ref{2Sm}), one can use the $c$ values above to give
\begin{align*}
(1/2,3/2)^{+}  &  :M(\Xi_{cc})_{1s}=3691.\allowbreak7+70.66\left\{
-1,1/2\right\}  =\left\{  3620.8,3726.8\right\}  \ \text{MeV,}\\
(1/2,3/2)^{+}  &  :M(\Omega_{cc})_{1s}=3789.5+72.3\left\{  -1,1/2\right\}
=\left\{  3717.\allowbreak2,3825.7\right\}  \ \text{MeV,}%
\end{align*}
where the mean mass(3691 MeV and 3789.5 MeV) in the 1s wave($v_{cc}^{2}=\allowbreak0.208$) are obtained for the systems $q\{cc\}$ and the two numbers in curly braces correspond to the respective states with $J^{P}=1/2^{+}$ and $3/2^{+}$.

Further, similar calculation applying to the 1s wave $\Xi_{bb}$ and $\Omega_{bb}$ gives
\begin{align*}
c\left(  \Xi_{bb}\right)  _{1s}  &  =c\left(  D\right)_{1s}\left(
\frac{M_{c}}{M_{bb}}\right)  =(140.6\ \text{MeV})\left(  \frac{1.44}%
{8.9167}\right)  =22.7\ \text{MeV,}\\
c\left(  \Omega_{bb}\right)  _{1s}  &  =c\left(  D_{s}\right)  _{1s}\left(
\frac{M_{c}}{M_{bb}}\right)  =(143.8\ \text{MeV})\left(  \frac{1.44}%
{8.9167}\right)  =23.2\ \text{MeV,}%
\end{align*}
and the ground state masses (with $v_{bb}^{2}=0.121$)
\begin{align*}
(1/2,3/2)^{+}  &  :M(\Xi_{bb})_{1s}=10\allowbreak226+22.7\left\{
-1,1/2\right\}  =\left\{  10203.3,10237.4\right\}  \ \text{MeV,}\\
(1/2,3/2)^{+}  &  :M(\Omega_{bb})_{1s}=10324+23.2\left\{  -1,1/2\right\}
=\left\{  10301.8,10335.6\right\}  \ \text{MeV,}%
\end{align*}
where $\bar{M}(q\{bb\})_{1S}=M_{bb}+m_{q}+M_{bb}v_{bb}^{2}=10\allowbreak
226.0$($10\allowbreak324.0$) MeV have been used for $q=n$(or $s$).

Finally, for the DH systems $q(bc)$, there are three ground states, $\Xi
_{bc}=$ $n\{bc\}$ with $J^{P}=1/2^{+}$ and $3/2^{+}$, $\Xi_{bc}^{\prime}=$
$n[bc]$ with $J^{P}=1/2^{+}$, and three ground states $\Omega_{bc}=$ $s\{bc\}$
with $J^{P}=1/2^{+}$ and $3/2^{+}$, $\Omega_{bc}^{\prime}=$ $s[bc]$ with
$J^{P}=1/2^{+}$. Notice that $v_{bc}^{2}=0.1428(0.1429)$ for the diquark $\{bc\}$($[bc]$), one finds, by Eq. (\ref{ReggS}), the mean masses for them to be%
\begin{align*}
\bar{M}(n\{bc\})_{1s}  &  =M_{\{bc\}}+m_{n}+M_{\{bc\}}v_{bc}^{2}%
=6964.3\ \text{MeV,}\\
\bar{M}(s\{bc\})_{1s}  &  =M_{\{bc\}}+m_{s}+M_{\{bc\}}v_{bc}^{2}%
=7062.3\ \text{MeV,}\\
\bar{M}(n[bc])_{1s}  &  =6963.0\ \text{MeV, }\bar{M}(s[bc])_{1S}%
=7061.0\ \text{MeV,}%
\end{align*}
and the $c$ parameter to be
\begin{align*}
c\left(  \Xi_{bc}\right)  _{1s}  &  =c\left(  D\right)  _{1s}\left(
\frac{M_{c}}{M_{\{bc\}}}\right)  =(140.6\ \text{MeV})\left(  \frac
{1.44}{5.9823}\right)  =33.8\ \text{MeV,}\\
c\left(  \Omega_{bc}\right)_{1s}  &  =c\left(  D_{s}\right)  _{1s}\left(
\frac{M_{c}}{M_{\{bc\}}}\right)  =(143.8\ \text{MeV})\left(  \frac
{1.44}{5.9823}\right)  =34.6\ \text{MeV.}%
\end{align*}

The ground state masses of the baryons $\Xi_{bc}$ and $\Omega_{bc}$ are then%
\begin{align*}
(1/2,3/2)^{+}  &  :M(\Xi_{bc})_{1s}=6964.3+33.8\left\{  -1,1/2\right\}
=\left\{  6930.5,6981.2\right\}  \ \text{MeV,}\\
(1/2,3/2)^{+}  &  :M(\Omega_{bc})_{1s}=7062.3+34.6\{-1,1/2\}=\left\{
7027.7,7079.\allowbreak6\right\}  \ \text{MeV,}\\
1/2^{+}  &  :M(\Xi_{bc}^{\prime})_{1s}=6963.0\ \text{MeV, }M(\Omega
_{bc}^{\prime})_{1s}=7061.0\ \text{MeV,}%
\end{align*}
with the two numbers in curly braces corresponding to $J^{P}=1/2^{+}$
and $3/2^{+}$, respectively. Our predictions for the ground-state masses of the most DH baryons ($\Omega_{Q\acute {Q}^{\prime}}$) are in consistent with that by Ref. \cite{EFG:D02}.

\subsection{Heavy pair binding and effective masses of excited diquarks }
We first explore and extract binding energies of heavy pairs in DH baryons and then compute the effective masses of heavy diquarks in its excited states. The later is done via estimating the energy shifts due to diquark excitations.

In QCD string picture, one simple and direct way to extract heavy pair binding is to use the "half" rule for the short-distance interaction of the $Q\bar{Q}^{\prime}$ systems and to subtract from the hadron masses of the $Q\bar{Q}^{\prime}$ systems (short string plus binding energy plus $2M_{Q}$) all involved (heavy and light) masses and string energies of mesons, leaving only the binding energy between $Q$ and $\bar{Q}^{\prime}$. The relation for the $Q\bar{Q}^{\prime}$ systems is
\begin{equation}
B(Q\bar{Q}^{\prime})_{N,L}=\bar{M}(Q\bar{Q}^{\prime})-\bar{M}(Q\bar{n}%
)-\bar{M}(Q\bar{n}^{\prime})+\bar{M}(n\bar{n}^{\prime}),\label{BQQ}%
\end{equation}
in which $\bar{M}$ represents the mean(spin-averaged) masses of the respective
mesons formed by the quark pairs $Q\bar{Q}^{\prime}$, $Q\bar{n}$ and $n\bar
{n}^{\prime}$($n=u,d$). The experimental mean-mass data \cite{ZylaPDG:D20} for these mesons are shown collectively in Table VIII, where the mean ($1S$) masses of the $c\bar{c}$ system, for instance,

\setlength{\abovecaptionskip}{0.4cm}
\begin{table}[h]
\caption{Means masses (MeV) of heavy quarkonia, $B_{c}$ mesons, the $B$
mesons, $D$ mesons and the light unflavored mesons. The data comes from the spin-averaging of the measured masses for these mesons \cite{ZylaPDG:D20}. }%
\begin{tabular}
[c]{ccccccc}\hline\hline
\textrm{Mesons} & \textrm{$c\bar{c}$} & \textrm{$b\bar{b}$} &\textrm{$b\bar{c}$} & \textrm{$c\bar{n}$} & \textrm{$b\bar{n}$} &\textrm{$n\bar{n}$}
\\\hline
$\bar{M}(1S)$ &3068.65  &9444.9  &6320.25  &1973.23  &5313.40  &619.98   \\
$\bar{M}(2S)$ &3627.50  &10017.3 &6901.20  &2627.00  &5917.00  &1423.75   \\
$\bar{M}(1P)$ &3525.26  &9899.7  &6765.32  &2435.72  &5737.17  &1245.79   \\\hline\hline
\end{tabular}
\end{table}
are $\bar{M}(c\bar{c})_{1S}=[3M(J/\psi)+M(\eta_{c})]/4=$ $3068.65$ MeV, $\bar{M}(c\bar{n})_{1S}=[3M(D^{\ast})+M(D)]/4=1973.23$ MeV, $\bar{M}(n\bar{n}^{\prime})_{1S}=[3M(\rho)+154]/4=619.98$ MeV.

Given the data in Table VIII, the binding energy of the heavy quark-antiquark is calculable by Eq. (\ref{BQQ}). In the case of $QQ^{\prime}=c\bar{c}$, Eq. (\ref{BQQ}) gives
\begin{align}
B(c\bar{c})_{1S} &  =\bar{M}(c\bar{c})_{1S}-2\bar{M}(c\bar{n})_{1S}+\bar
{M}(n\bar{n}^{\prime})_{1S},\nonumber\\
&  =3068.65-2(1973.23)+619.98,\nonumber\\
&  =-257.\,\allowbreak83\text{ MeV, }\label{B1S}%
\end{align}
where we have used the computed 1S-wave mass $154.0$ MeV of the pion in Ref. \cite{EFG:D09}, instead of the observed (very light) mass of the physical pion, for which an extra suppression mechanism (Nambu-Goldstone mechanism) enters due to the breaking of chiral symmetry. Note that the Nambu-Goldstone mechanism for the heavy quarkonia is ignorable and not comparable with that of the physical pion. With the 1S wave data of the respective bottomonium, the $B$ meson and
$\bar{M}(n\bar{n}^{\prime})_{1S}=619.98$ MeV in Table VIII, one can similarly estimate $B(b\bar{b})$ via Eq. (\ref{BQQ}). The results are listed in Table IX.

For the pair $b\bar{c}$, only the 1S and 2S wave mesons(the $B_{c}$) are
available for $\bar{M}(b\bar{c})$ experimentally, and they give, by Eq.
(\ref{BQQ}),
\begin{align*}
B(b\bar{c})_{1S} &  =\bar{M}(b\bar{c})_{1S}-\bar{M}(b\bar{n})_{1S}-\bar
{M}(\bar{c}n)_{1S}+\bar{M}(n\bar{n})_{1S},\\
&  =6320.3-5313.4-1973.2+619.98,\\
&  =-346.4\text{ MeV, }%
\end{align*}
in which $\bar{M}(b\bar{c})_{1S}$ is estimated by the mean mass of the measured $M(B_{c},0^{-})_{1S}=6274.5$ MeV and the predicted mass splitting $\Delta
M(B_{c})_{1S}=61$ MeV between the $1^{-}$ and $0^{-}$ states by Ref.
\cite{EFG:D11}:
\[
\bar{M}(b\bar{c})_{1S}=\frac{1}{4}[6274.5+3\times(6274.5+\Delta M(B_{c}%
)_{1S})]=6320.3\text{ MeV.}%
\]
$\qquad$
For 2S wave $b\bar{c}$, one can use the measured 2S-wave mass $M(B_{c}%
,0^{-})_{2S}=6871.2$ MeV of the $B_{c}$ meson and the predicted mass splitting
$\Delta M(B_{c})_{2S}=40$ MeV by Ref. \cite{EFG:D11} to get
\[
\bar{M}(b\bar{c})_{2S}=\frac{1}{4}[6871.2+3\times(6871.2+\Delta M(B_{c}%
)_{2S})]=6901.2\text{ MeV,}%
\]
and thereby to find
\begin{align*}
B(b\bar{c})_{2S} &  =6901.2-5917.0-2607.5+1423.7=-199.6\text{ MeV,}
\end{align*}
where $\bar{M}(c\bar{n})_{2S}=2607.5$ MeV\cite{ZylaPDG:D20}, and $\bar
{M}(b\bar{n})_{2S}=5917.0$ MeV, taken from the predicted mean
mass\cite{JiaD:E19} of the $B$ mesons in 2S-wave. Similarly, one can find $B(c\bar{c})_{2S}=-117.3$ MeV and $B(c\bar{c})_{2S}=-392.98$ MeV, as shown in Table IX.

\vspace{0cm}
\setlength{\abovecaptionskip}{0cm}
\begin{table}[h]
\caption{The binding energy$B(Q\bar{Q}^{\prime})$, $B(QQ^{\prime})$ and their shift $\Delta B$ relative to the ground state, computed from Eq. (\ref{BQQ}), Eq. (\ref{Half}) and Eq. (\ref{dBB}), respectively. All items are in MeV. }%
\begin{tabular}
[c]{cccccc}\hline\hline
\textrm{Binding energy} & \textrm{$1S$} & \textrm{$2S$} &
\textrm{$1P$} & $\Delta$\textrm{$B(2S)$} & $\Delta B(1P)$\\\hline
$B(c\bar{c})$ &-257.83  &-117.30  &-100.38  &140.53  &157.45   \\
$B(b\bar{b})$ &-561.92  &-392.98  &-328.81  &168.94  &233.11   \\
$B(b\bar{c})$ &-346.40  &-199.55  &-161.78  &146.85  &184.32   \\
$B(cc)$       &-128.92  &-58.65   &-50.19   &70.27   &78.73   \\
$B(bb)$       &-280.96  &-196.49  &-164.41  &84.47   &116.55   \\
$B(bc)$       &-173.20  &-99.78   &-80.89   &73.43   &92.32   \\\hline\hline
\end{tabular}
\end{table}

For the $B_{c}$ meson, no data in P wave is available. We estimate it via interpolating the P wave masses of the $c\bar{c}$ and $b\bar{b}$ systems which are known experimentally. Inspired by atomic spectra in a purely Coulombic potential, we write the binding energy $B(Q\bar{Q}^{\prime})$ as a power form of the reduced pair mass $\mu_{QQ^{\prime}}=M_{Q}M_{Q^{\prime}}/(M_{Q}+M_{Q^{\prime}})$
\cite{KR:D18,ZXJ:D21},
\begin{equation}
B(Q\bar{Q}^{\prime})-B_{0}=k[\mu_{QQ^{\prime}}]^{P}=k\left(  \frac
{M_{Q}M_{Q^{\prime}}}{M_{Q}+M_{Q^{\prime}}}\right)  ^{P}.\label{dBNL}%
\end{equation}
where $B_{0}$ are a constant while the parameters $k=k_{N,L}$ and
$P=P_{N,L}$ depend on the radial and angular quantum numbers of the excited
$Q\bar{Q}^{\prime}$. In the ground($1S1s$) state, the binding
$B(c\bar{c})=-257.83$ MeV, $B(b\bar{b})=-561.92$ MeV$\ $and $B(b\bar
{c})=-346.40$ MeV in Table IX correspond, by Eq. (\ref{dBNL}), to the parameters \{$B_{0}=62.\allowbreak066$ MeV, $P_{0,0}=0.58805$, $k_{0,0}=-388.342$ \}.

When $Q\bar{Q}^{\prime}$ excited to the 1P wave, substitution of the measured
mean masses of the charmonium and bottomonium in Table VIII into Eq. (\ref{BQQ}) leads to,
\[
B(c\bar{c})_{1P}=-100.38\text{ MeV, }B(b\bar{b})_{1P}=-328.81\text{ MeV.}%
\]
One can then interpolate these two binding energies via Eq. (\ref{dBNL}) to predict (for the $b\bar{c}$ system) $P_{0,1}=0.77365$, $k_{0,1}=-0.209448$, and further gives, by Eq. (\ref{dBNL}),
\begin{align}
B(b\bar{c})_{1P}  &  =B_{0}+\frac{B(c\bar{c})_{1P}-B_{0}}%
{[(1+1.44/4.48)/2]^{P_{0,1}}},\nonumber\\
&  =62.\allowbreak066+\frac{-100.377-(62.\allowbreak066)}%
{[(1+1.44/4.48)/2]^{0.773645}}\nonumber\\
&  =-161.78\text{ MeV.} \label{Bbcb}%
\end{align}
as shown in Table IX.

For the heavy quark-antiquark pairs $\{c\bar{c},b\bar{b},b\bar{c}\}$, three
values of string tension $T$ that reproduce, by Eq. (\ref{Mbnl}), the mean
masses in Table VIII and the binding energies in Table IX, are given by
\begin{equation}
\{T(c\bar{c}),T(b\bar{b}),T(b\bar{c})\}=\left\{
0.21891,0.43367,0.34278\right\}  \text{ GeV}^{2}\text{.} \label{TQQ}%
\end{equation}

Next, we consider binding energy $B(QQ^{\prime})$ of heavy quark pair
$QQ^{\prime}$ in the color antitriplet($\bar{3}_{c}$). In a baryon $qQQ^{\prime}$, such a binding enters as a correction to
energy of the QCD string connecting $Q$ and $Q^{\prime}$ in
short-distance, as indicated by heavy-quarkonia spectra\cite{ZylaPDG:D20}.
By color-$SU(3)$ argument and by lattice simulations in Ref. \cite{NB09}, the $QQ^{\prime}$ interaction strength in a color triplet is half of that of the $Q\bar{Q}^{\prime}$ in a color singlet when the interquark(QQ) distance is small. One can then write, as in Ref. \cite{KR:D14},
\begin{equation}
B(QQ^{\prime})=\frac{1}{2}B(Q\bar{Q}^{\prime}), \label{Half}%
\end{equation}
with $Q\bar{Q}^{\prime}$ in color singlet($1_{c}$). Relative to the ground states($1S1s$), the shift $\Delta B_{N,L}\equiv B(QQ^{\prime})_{N,L}-B(QQ^{\prime})_{0,0}$ of the binding energy are, by Eq. (\ref{Half}),
\begin{equation}
\Delta B(QQ)_{N,L}=\frac{1}{2}[B(Q\bar{Q}^{\prime})_{N,L}-B(Q\bar{Q}^{\prime
})_{0,0}]. \label{dBB}%
\end{equation}

Using the binding data for the $Q\bar{Q}^{\prime}$ in Table IX, Eq. (\ref{Half}) and Eq. (\ref{dBB}) give all binding shifts $\Delta B(Q\bar{Q})$ for the 2S and 1P states, and all binding shifts $\Delta B(QQ^{\prime})$ for the 2S and 1P states, as listed collectively in Table IX.

With the binding data in Table IX, and the values of $T$ in Eq. (\ref{TQQ}), Eq. (\ref{dMd}) gives rise to the mean-mass shifts
of the DH baryons due to the (2S and 1P) diquark excitations relative to their
ground states($1S1s$),
\begin{align}
\{\Delta\bar{M}^{(\ast)}(\Xi_{cc}), \Delta\bar{M}^{(\ast)}(\Xi_{bb}) , \Delta\bar{M}^{(\ast)}(\Xi_{bc}) \}_{2S} &
=\{393.56,347.02,365.97\}\text{ MeV,}\label{dMb2}\\
\{\Delta\bar{M}^{(\ast)}(\Xi_{cc}), \Delta\bar{M}^{(\ast)}(\Xi_{bb}) , \Delta\bar{M}^{(\ast)}(\Xi_{bc}) \}_{1P} &
=\{430.87,404.39,412.27\}\text{ MeV.}\label{dMb1}%
\end{align}
and the same values for the $\Omega_{cc^{\prime}}$, $\Omega_{bb^{\prime}}$ and $\Omega_{bc^{\prime}}$. This enables us to define effective masses $E_{QQ^{\prime}}$ of heavy-diquark
in its excited state via the energy shift due to diquark excitations:
\[
(E_{QQ^{\prime}})_{N,L}=M_{QQ^{\prime}}+(\Delta\bar{M}^{(\ast)})_{N,L},
\]
which gives explicitly, with the usage of Eq. (\ref{hdiquark}), Eq. (\ref{Hdiq}), Eq. (\ref{dMb2}) and Eq. (\ref{dMb1}),
\begin{align}
(E_{cc},E_{bb},E_{bc})_{2S} &  =\{3.2591,9.2637,6.2583\}\text{ GeV,}\label{Ecc2S}\\
(E_{cc},E_{bb},E_{bc})_{1P} &  =\{3.2964,9.3211,6.3046\}\text{ GeV.}\label{Ecc1P}
\end{align}

\subsection{Spin couplings due to diquark excitations}
Let us consider the $2S1s$ and $1P1s$ states of the DH baryons. For the spin couplings of the baryon multiplets, we utilize the mass scaling relative to the ground state $D$ mesons (the $D^{\pm}$ with mass
$1869.66\pm0.05$ MeV and the $D^{\ast}(2010)^{\pm}$ with mass $2010.26\pm0.05$
MeV\cite{ZylaPDG:D20}),
\begin{equation}
c^{\ast}\left(  \Xi_{bb}(bbq)\right)  _{2S,1P}=\left(  \frac{M_{c}}%
{(E_{bb})_{2S,1P}}\right)  \left(  \frac{m_{n}}{m_{q}}\right)  c\left(
D\right)  _{1S},\label{cMS}%
\end{equation}
in which $c\left(D\right)_{1S}=M(D^{\ast\pm})-M\left(  D^{\pm}\right)
=140.6$ MeV (that corresponds to the contact-term) for the $D$ mesons, as Ref. \cite{JiaLH:19}.

(1) The $2S1s$ states. Using data in Eq. (\ref{dMb2}) and Eq. (\ref{dMb1}), Eq. (\ref{cMS}) leads to
\begin{align}
c^{\ast}\left(  \Xi_{bb}(bbn)\right)  _{2S} &  =\left(  \frac{M_{c}}{E_{bb}%
}\right)  (140.6\text{ MeV})=\allowbreak21.86\text{ MeV,}\nonumber\\
c^{\ast}\left(  \Omega_{bb}(bbs)\right)  _{2S} &  =\left(  \frac{M_{c}}%
{E_{bb}}\right)  \left(  \frac{m_{n}}{m_{s}}\right)  (140.6\text{
MeV})=\allowbreak15.33\text{ MeV,}\label{cbbn}%
\end{align}
where $(E_{bb})_{2S}=9.2637$ GeV.
Further application of the following scaling relations between the doubly charmed and bottome baryons%
\begin{align}
c^{\ast}\left(  \Xi_{cc}\right)  _{2S} &  =c^{\ast}\left(  \Xi_{bb}\right)
\left(  \frac{E_{bb}}{E_{cc}}\right)  ,c^{\ast}\left(  \Xi_{bc}\right)
_{2S}=c^{\ast}\left(  \Xi_{bb}\right)  \left(  \frac{E_{bb}}{E_{bc}}\right)
,\nonumber\\
c^{\ast}\left(  \Omega_{cc}\right)  _{2S} &  =c^{\ast}\left(  \Omega
_{bb}\right)  \left(  \frac{E_{bb}}{E_{cc}}\right)  ,c^{\ast}\left(
\Omega_{bc}\right)  _{2S}=c^{\ast}\left(  \Omega_{bb}\right)  \left(
\frac{E_{bb}}{E_{bc}}\right)  ,\label{cSR}%
\end{align}
gives
\begin{align}
c^{\ast}\left(  \Xi_{cc}\right)  _{2S} &  =62.\allowbreak12\text{ MeV,
}c^{\ast}\left(  \Xi_{bc}\right)  _{2S}=32.35\text{ MeV, }\nonumber\\
c^{\ast}\left(  \Omega_{cc}\right)  _{2S} &  =43.56\text{ MeV, }c^{\ast
}\left(  \Omega_{bc}\right)  _{2S}=22.69\text{ MeV.}\label{c2S}%
\end{align}

Given the masses in Table II and the mean masses of the ground-state baryons
$\Omega_{cc}$, $\Omega_{bb}$ and $\Omega_{bc}$ in Ref. \cite{EFG:D02}, $\Delta\bar{M}^{(\ast)}$ in Eq. (\ref{dMb2}) and
the values of $c^{\ast}$ in Eqs. (\ref{cbbn}) and (\ref{cSR}), one gets, by
Eq. (\ref{SSM}), the multiplet masses of the DH baryons $\Xi_{QQ^{\prime}}$
and $\Omega_{QQ^{\prime}}$ in the $2S1s$ state, listed in Table X through XVI, and shown in FIGs 1-8, respectively.

(2) The $1P1s$ states.
Application of the mass scaling Eq. (\ref{cMS}) leads to($E_{bb}=9.3211$ GeV)%
\begin{align}
c^{\ast}\left(  \Xi_{bb}\right)  _{1P}  &  =\left(  \frac{M_{c}}{E_{bb}%
}\right)  (140.6\text{ MeV})=\allowbreak21.\allowbreak72\text{ MeV,}%
\nonumber\\
c^{\ast}\left(  \Omega_{bb}\right)  _{1P}  &  =\left(  \frac{M_{c}}{E_{bb}%
}\right)  \left(  \frac{m_{n}}{m_{s}}\right)  (140.6\text{ MeV})=\allowbreak
15.23\text{ MeV,} \label{csbb}%
\end{align}
and
\begin{align}
c^{\ast}\left(  \Xi_{cc}\right)  _{1P}  &  =c^{\ast}\left(  \Xi_{bb}\right)
_{1P}\left(  \frac{9.3211}{3.2964}\right)  =61.42\text{ MeV,}\nonumber\\
c^{\ast}\left(  \Xi_{bc}\right)  _{1P}  &  =c^{\ast}\left(  \Xi_{bb}\right)
_{1P}\left(  \frac{9.3211}{6.3046}\right)  =32.31\text{ MeV,}\nonumber\\
c^{\ast}\left(  \Omega_{cc}\right)  _{1P}  &  =c^{\ast}\left(  \Omega
_{bb}\right)  _{1P}\left(  \frac{9.3211}{3.2964}\right)  =43.07\text{
MeV,}\nonumber\\
c^{\ast}\left(  \Omega_{bc}\right)  _{1P}  &  =c^{\ast}\left(  \Omega
_{bb}\right)  _{1P}\left(  \frac{9.3211}{6.3046}\right)  =22.66\text{ MeV,}
\label{csbc}%
\end{align}

Given the masses in Table II and the mean masses of the ground-state baryons
$\Omega_{cc}$, $\Omega_{bb}$ and $\Omega_{bc}$ in Ref. \cite{EFG:D02},
$\Delta\bar{M}^{(\ast)}$ in Eq. (\ref{dMb1}) and the $c^{\ast}$ values in Eqs.
(\ref{csbb}) and (\ref{csbc}), one can apply Eq. (\ref{SSM1}) and Eq.
(\ref{LssM}) to obtain the mean mass $\bar{M}_{1P}+(\Delta\bar{M}^{(\ast
)})_{1P}$ and the multiplet masses of the $1P1s$-wave DH baryons. The results are listed in Table X through XVI and shown in FIGs. 1-8, respectively, for the $2S1s$ and $1P1s$ states of the DH baryons.

We collect all obtained masses of the DH baryons, which are computed via summing of the spin averaged masses($\bar{M}$) in section III and the mass splitting $\Delta M$ in section IV, in Tables X through XVI. The ground state masses obtained in Section VI are also listed there.

\renewcommand{\tabcolsep}{0.3cm}
\vspace{0.2 cm}
\begin{table}[h]
\caption{ Mass spectra of the baryon $\Xi_{cc}$ (in MeV). Our results are obtained by the respective sum of the mean mass of the $\Xi_{cc}$ in Table III, the binding energy shift(vanishes in 1S1s wave) in Table IX and its splittings in Section III with the couplings given in Section IV. }%
\begin{tabular}
[c]{cccccccccc}\hline\hline
 $%
\begin{array}
[c]{c}%
\text{$\left(N L n_{q} l\right) J^{P}$}%
\end{array}
$ & This work & \cite{EFG:D02} & \cite{Lu:2017meb} & \cite{Gershtein:2000nx} &
\cite{SotoCast:D21} & \cite{Yoshida:2015tia}& \cite{PEMP} & \cite{Eakins:2012jk}& \cite{Roberts:2007ni}\\\hline
         $(1S1p)1/2^-$ &3970.3 &4053 &3998 &3927 &$4030$&3947 &4064&4081 &3910\\
 $(1S1p)1/2^{\prime-}$ &4082.5 &4136 &3985 &4052 &$4070$&4135 &4174&4073 &4074\\
         $(1S1p)3/2^-$ &4039.9 &4101 &4014 &4039 &$4080$&3949 &4075&4077 &3921\\
 $(1S1p)3/2^{\prime-}$ &4100.6 &4196 &4025 &4034 &$4120$&4137 &4164&4079 &4078\\
         $(1S1p)5/2^-$ &4130.9 &4155 &4050 &4047 &$4075$&4163 &4194&4089 &4092 \\\hline
         $(1S2s)1/2^+$ &4216.3 & $-$ &4172 & $-$ &4370&4159 &4304&4311 &$-$ \\
         $(1S2s)3/2^+$ &4275.1 & $-$ &4193 & $-$ &4330&4131 &4264&4368 &$-$ \\\hline
         $(2S1s)1/2^+$ &4023.1 &3910 &4004 &3812 &$4270$ &4079 &$4314$&4030 &4029\\
         $(2S1s)3/2^+$ &4116.3 &4027 &4036 &3944 &$4290$ &4114 &$4315$ &4078 &4042\\\hline
         $(1P1s)1/2^-$ &4061.1 &3838 &3873 &3702 &$4530$ &4149 &$4024$ &3911&$-$ \\
         $(1P1s)3/2^-$ &4153.3 &3959 &3916 &3834 &$4560$ &4159 &$4194$ &3917 &$-$ \\\hline\hline
\end{tabular}
\end{table}

\renewcommand{\tabcolsep}{0.3cm}
\setlength{\abovecaptionskip}{0.1cm}
\begin{table}[h]
\caption{ Mass spectra of $\Xi_{bb}$ baryons (in MeV). Our results are obtained by the respective sum of the mean mass of the $\Xi_{bb}$ in Table III, the binding energy shift(vanishes in 1S1s wave) in Table IX and its splittings in Section III with the couplings given in Section IV. }%
\begin{tabular}
[c]{cccccccccc}\hline\hline
 $%
\begin{array}
[c]{c}%
\text{$\left(N L n_{q} l\right) J^{P}$}%
\end{array}
$ & This work & \cite{EFG:D02} & \cite{Lu:2017meb} & \cite{Gershtein:2000nx} &
\cite{SotoCast:D21} & \cite{Yoshida:2015tia}& \cite{Shah:2017liu} & \cite{Eakins:2012jk}& \cite{Roberts:2007ni}\\\hline
         $(1S1p)1/2^-$ &10523.1 &10632 &10525 &10541 &$10380$&10476&10511&10694 &10493\\
 $(1S1p)1/2^{\prime-}$ &10559.0 &10675 &10504 &10578 &$10520$&$-$  &10514&10691 &$-$ \\
         $(1S1p)3/2^-$ &10545.3 &10647 &10526 &10567 &$10410$&10476&10506&10691 &10495 \\
 $(1S1p)3/2^{\prime-}$ &10564.8 &10694 &10528 &10581 &$10530$&$-$  &10509&10692 &$-$ \\
         $(1S1p)5/2^-$ &10574.5 &10661 &10547 &10580 &$10510$&10759&10518&10695 &10713 \\\hline
         $(1S2s)1/2^+$ &10707.9 &10832 &10662 & $-$ &10580&10612&10609&10940 &$-$ \\
         $(1S2s)3/2^+$ &10726.8 &10860 &10675 & $-$ &10570&10593&10617&10972 &$-$ \\\hline
         $(2S1s)1/2^+$ &10550.5 &10441 &10464 &10373 &$10530$ &10571 &$-$ &10551 &10571 \\
         $(2S1s)3/2^+$ &10583.2 &10482 &10480 &10413 &$10550$ &10592 &$-$ &10574 &10592\\\hline
         $(1P1s)1/2^-$ &10608.0 &10368 &10364 &10310 &$10720$ &10740 &$-$ &10470 &$-$  \\
         $(1P1s)3/2^-$ &10640.6 &10408 &10387 &10343 &$10740$ &10742 &$-$ &10470 &$-$ \\\hline\hline
\end{tabular}
\end{table}

\setlength{\abovecaptionskip}{0.4cm}
\begin{table}[h]
\caption{ Mass spectra(MeV) of $\Xi_{bc}$ baryons. Our results are obtained by the respective sum of the mean mass of the $\Xi_{bc}$ in Table III, the binding energy shift(vanishes in 1S1s wave) in Table IX and its splittings in Section III with the couplings given in Section IV. }%
\begin{tabular}
[c]{ccccccc}\hline\hline
 $%
\begin{array}
[c]{c}%
\text{State}\\
\text{$\left(N L n_{q} l\right) J^{P}$}%
\end{array}
$ & This work & \cite{Giannuzzi:2009gh} & \cite{Shah:2017liu} & \cite{Eakins:2012jk} &
\cite{Mohajery:2018qhz} & \cite{Roberts:2007ni}\\\hline
         $(1S1p)1/2^-$ &7273.2 &$-$ &7156 &7397 &6842&7206\\
 $(1S1p)1/2^{\prime-}$ &7327.7 &$-$ &7161 &7390 &6847&7231   \\
         $(1S1p)3/2^-$ &7307.0 &$-$ &7144 &7392 &6831&7208 \\
 $(1S1p)3/2^{\prime-}$ &7336.6 &$-$ &7150 &7394 &6837&7229   \\
         $(1S1p)5/2^-$ &7351.3 &$-$ &7171 &7399 &6856&7272 \\\hline
         $(1S2s)1/2^+$ &7478.6 &7478 &7240 &7634 &6919&$-$ \\
         $(1S2s)3/2^+$ &7507.2 &7495 &7263 &7676 &6939&$-$ \\\hline
         $(2S1s)1/2^+$ &7297.9 &$-$ &$-$ &7321 &$-$ &7044  \\
         $(2S1s)3/2^+$ &7346.4 &$-$ &$-$ &7353 &$-$ &7386 \\\hline
         $(1P1s)1/2^-$ &7344.3 &$-$ &$-$ &7212 &$-$ &$-$  \\
         $(1P1s)3/2^-$ &7392.7 &$-$ &$-$ &7214 &$-$ &$-$  \\\hline\hline
\end{tabular}
\end{table}

\setlength{\abovecaptionskip}{0.4cm}
\begin{table}[h]
\caption{ Mass spectra of $\Omega_{cc}$ baryons(in MeV). Our results are obtained by the respective sum of the mean mass of the $\Omega_{cc}$ in Table III, the binding energy shift(vanishes in 1S1s wave) in Table IX and its splittings in Section III with the couplings given in Section IV. }%
\begin{tabular}
[c]{cccccccccc}\hline\hline
 $%
\begin{array}
[c]{c}%
\text{$\left(N_{d} L n_{q} l\right) J^{P}$}%
\end{array}
$ & This work & \cite{EFG:D02} & \cite{Lu:2017meb} & \cite{PEMP} &
\cite{Yoshida:2015tia} & \cite{Roberts:2007ni}& \cite{Shah:2016vmd} & \cite{Salehi:2019kvb}& \cite{Kiselev:2002iy}\\\hline
         $(1S1p)1/2^-$ &4102.0 &4208 &4087 &$4084$ &4086&4046 &3989&3965 &4050\\
 $(1S1p)1/2^{\prime-}$ &4184.1 &4271 &4081 &$4194$ &4199&4135 &3998&3987 &4145 \\
         $(1S1p)3/2^-$ &4165.2 &4252 &4107 &$4114$ &4086&4052 &3972&3956 &4176 \\
 $(1S1p)3/2^{\prime-}$ &4200.6 &4325 &4114 &$4164$ &4201&4140 &3981&3978 &4102 \\
         $(1S1p)5/2^-$ &4227.1 &4330 &4134 &$4264$ &4220&4152 &3958&3926 &4134 \\\hline
         $(1S2s)1/2^+$ &4343.2 &$-$  &4270 &4364&4295&4180 &4041&3964 &$-$ \\
         $(1S2s)3/2^+$ &4384.4 &$-$  &4288 &4354&4265&4188 &4096&3979 &$-$ \\\hline
         $(2S1s)1/2^+$ &4190.7 &4075 &4118 &$4314$ &4227 &$-$  &$-$ &$-$ &3925 \\
         $(2S1s)3/2^+$ &4256.1 &4174 &4142 &$4334$ &4263 &$-$  &$-$ &$-$ &4064\\\hline
         $(1P1s)1/2^-$ &4228.5 &4002 &3986 &$4074$ &4210 &$-$ &$-$ &$-$ &3812  \\
         $(1P1s)3/2^-$ &4293.1 &4102 &4020 &$4204$ &4218 &$-$ &$-$ &$-$ &3949 \\\hline\hline
\end{tabular}
\end{table}

\begin{table}[h]
\caption{ Mass spectra of $\Omega_{bb}$ baryons (in MeV). Our results are obtained by the respective sum of the mean mass of the $\Omega_{bb}$ in Table III, the binding energy shift(vanishes in 1S1s wave) in Table IX and its splittings in Section III with the couplings given in Section IV.  }%
\begin{tabular}
[c]{cccccccccc}\hline\hline
 $%
\begin{array}
[c]{c}
\text{$\left(N_{d} L n_{q} l\right) J^{P}$}%
\end{array}
$ & This work & \cite{EFG:D02} & \cite{Lu:2017meb} & \cite{Giannuzzi:2009gh} &
\cite{Yoshida:2015tia} & \cite{Roberts:2007ni}& \cite{Shah:2016vmd} & \cite{Salehi:2019kvb}& \cite{Kiselev:2002iy}\\\hline
         $(1S1p)1/2^-$ &10651.8 &10771 &10605 &$-$ &10607&10616 &10646&10968 &10651\\
 $(1S1p)1/2^{\prime-}$ &10678.2 &10804 &10591 &$-$ &10796&10763 &10648&10976 &10700 \\
         $(1S1p)3/2^-$ &10672.1 &10785 &10610 &$-$ &10608&10619 &10641&10957 &10661 \\
 $(1S1p)3/2^{\prime-}$ &10683.5 &10802 &10611 &$-$ &10797&10765 &10643&10963 &10720 \\
         $(1S1p)5/2^-$ &10692.3 &10798 &10625 &$-$ &10808&10766 &10637&10956 &10670 \\\hline
         $(1S2s)1/2^+$ &10845.2 &10970 &10751 &10830&10744&10693&10736&10969 &$-$ \\
         $(1S2s)3/2^+$ &10858.4 &10992 &10763 &10839&10730&10721&10617&10964 &$-$ \\\hline
         $(2S1s)1/2^+$ &10710.7 &10610 &10566 &$-$ &10707 &$-$  &$-$ &$-$ &10493 \\
         $(2S1s)3/2^+$ &10733.7 &10645 &10579 &$-$ &10723 &$-$  &$-$ &$-$ &10540\\\hline
         $(1P1s)1/2^-$ &10768.2 &10532 &10464 &$-$ &10803 &$-$ &$-$ &$-$ &10416  \\
         $(1P1s)3/2^-$ &10791.0 &10566 &10482 &$-$ &10805 &$-$ &$-$ &$-$ &10462 \\\hline\hline
\end{tabular}
\end{table}

\setlength{\abovecaptionskip}{0.3cm}
\begin{table}[h]
\caption{ Mass spectra of $\Omega_{bc}$ baryons (in MeV). Our results are obtained by the respective sum of the mean mass of the $\Omega_{bc}$ in Table III, the binding energy shift(vanishes in 1S1s wave) in Table IX and its splittings in Section III with the couplings given in Section IV. }%
\begin{tabular}
[c]{cccccc}\hline\hline
 $%
\begin{array}
[c]{c}%
\text{$\left(N_{d} L n_{q} l\right) J^{P}$}%
\end{array}
$ & This work & \cite{Giannuzzi:2009gh} &\cite{Roberts:2007ni}& \cite{Shah:2016vmd} & \cite{Salehi:2019kvb}\\\hline
         $(1S1p)1/2^-$ &7401.8 &$-$  &7335&7386 &7476\\
 $(1S1p)1/2^{\prime-}$ &7441.7 &$-$  &7346&7392 &7490   \\
         $(1S1p)3/2^-$ &7432.5 &$-$  &7334&7373 &7470 \\
 $(1S1p)3/2^{\prime-}$ &7449.7 &$-$  &7349&7379 &7486  \\
         $(1S1p)5/2^-$ &7463.0 &$-$  &7362&7363 &7451\\\hline
         $(1S2s)1/2^+$ &7611.5 &7559 &$-$ &7480 &7475\\
         $(1S2s)3/2^+$ &7631.7 &7571 &$-$ &7497 &7485\\\hline
         $(2S1s)1/2^+$ &7459.3 &$-$  &$-$ &$-$  &$-$\\
         $(2S1s)3/2^+$ &7493.3 &$-$  &$-$ &$-$  &$-$\\\hline
         $(1P1s)1/2^-$ &7505.6 &$-$  &$-$ &$-$  &$-$\\
         $(1P1s)3/2^-$ &7539.8 &$-$
         &$-$ &$-$ &$-$ \\\hline\hline
\end{tabular}
\end{table}

\setlength{\abovecaptionskip}{0.4cm}
\begin{table}[h]
\caption{ Mass spectra of $\Xi_{bc}^{\prime}/\Omega_{bc}^{\prime}$ baryons (in MeV). Our results are obtained by the respective sum of the mean mass of the $\Xi_{bc}^{\prime}/\Omega_{bc}^{\prime}$ in Table III, the binding energy shift(vanishes in 1S1s wave) in Table IX and its splittings in Section III with the couplings given in Section IV.  }%
\begin{tabular}
[c]{cccc}\hline\hline
 $%
\begin{array}
[c]{c}%
\text{$\left(N_{d} L n_{q} l\right) J^{P}$}%
\end{array}
$ & This work($\Xi_{bc}^{\prime}$) & \cite{Eakins:2012jk} & This work($\Omega_{bc}^{\prime}$) \\\hline
         $(1S1p)1/2^-$ &7294.7 &7388  &7421.2  \\
         $(1S1p)3/2^-$ &7341.4 &7390  &7453.9   \\\hline
         $(1S2s)1/2^+$ &7497.0 &7645  &7624.2  \\
         $(2S1s)1/2^+$ &7329.0 &7333  &7528.3   \\\hline
         $(1P1s)1/2^-$ &7375.3 &7230  &7528.3   \\
 $(1P1s)1/2^{\prime-}$ &7343.0 &7199  &7505.6     \\
         $(1P1s)3/2^-$ &7391.4 &7228  &7539.6   \\
 $(1P1s)3/2^{\prime-}$ &7326.8 &7201  &7494.3     \\
         $(1P1s)5/2^-$ &7407.6 &7265  &7550.9   \\\hline\hline
\end{tabular}
\end{table}

\setlength{\abovecaptionskip}{-0.5cm}
\begin{figure}[!ht]
\begin{center}
\includegraphics[width=0.6\textwidth]{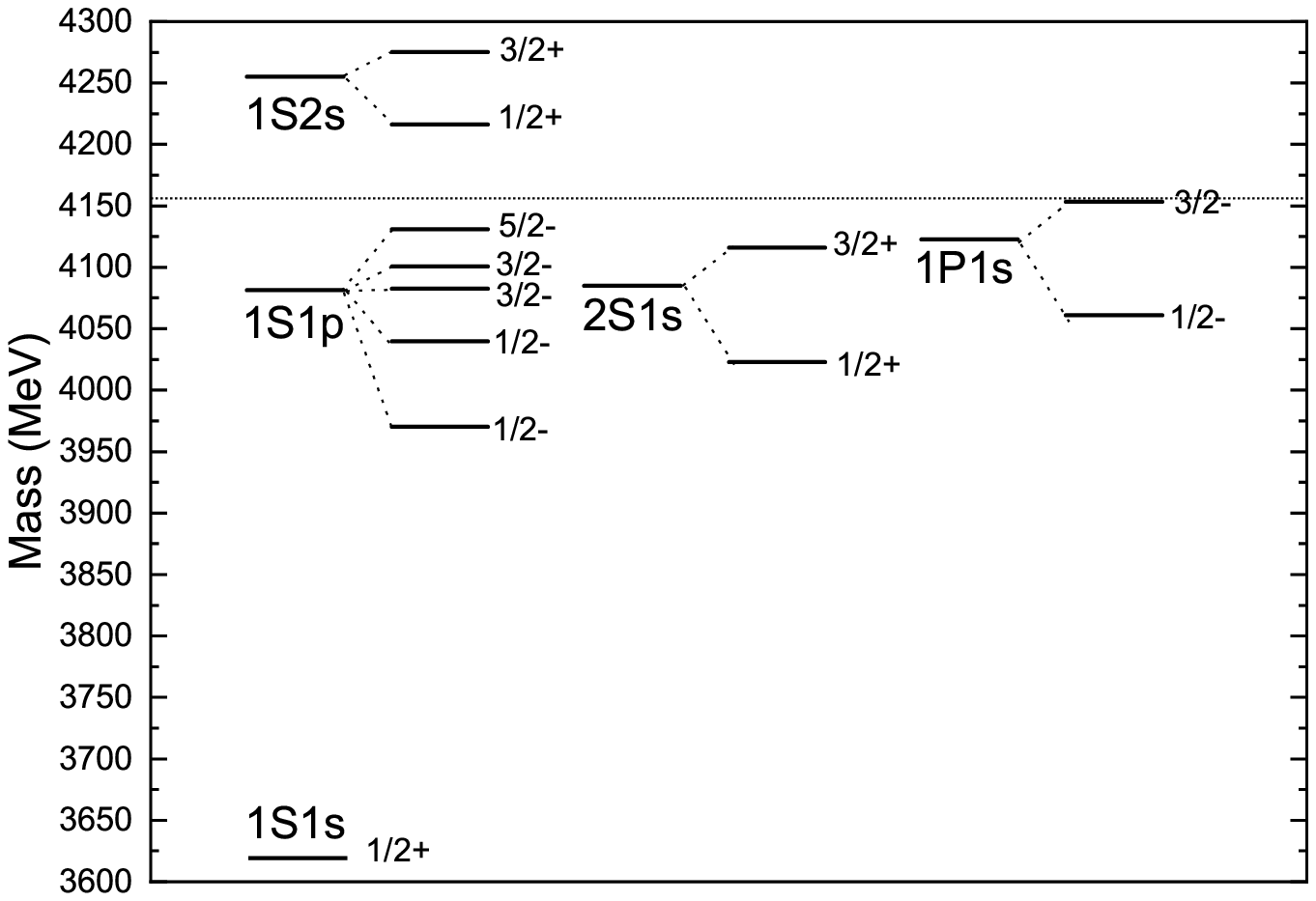}
\end{center}
\caption{Mass spectrum(in MeV) of the $\Xi_{cc}$ baryons, corresponding to the Table X. The horizontal dashed line shows the $\Lambda_{c}D$ threshold.}
\label{fig:Xicc}
\end{figure}
\setlength{\abovecaptionskip}{-0.4cm}
\begin{figure}[!ht]
\begin{center}
\includegraphics[width=0.6\textwidth]{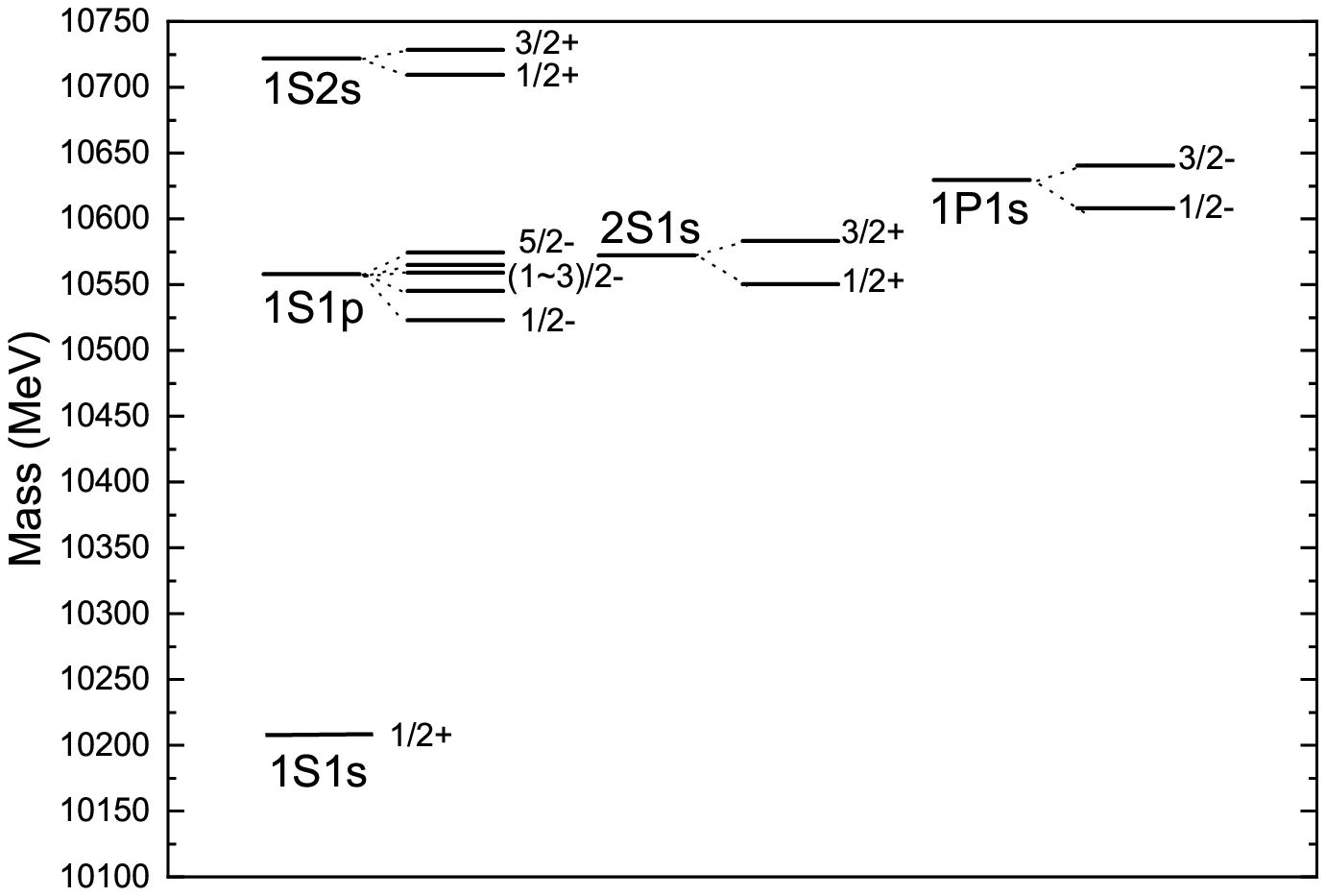}
\end{center}
\caption{Mass spectrum(in MeV) of the $\Xi_{bb}$ baryons, corresponding to the Table XI.}
\label{fig:Xibb}
\end{figure}

\setlength{\abovecaptionskip}{-0.4cm}
\begin{figure}[!ht]
\begin{center}
\includegraphics[width=0.6\textwidth]{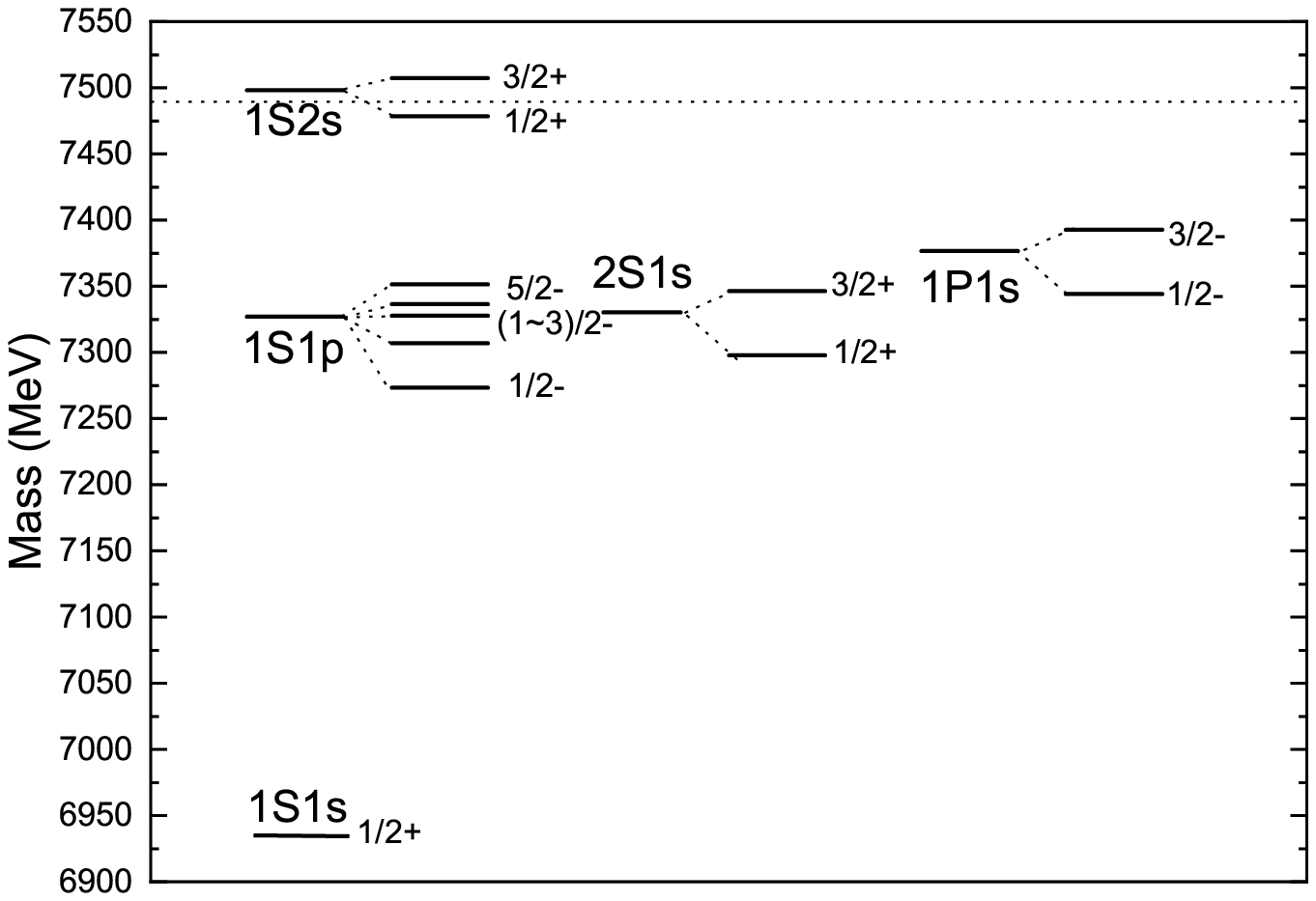}
\end{center}
\caption{Mass spectrum(in MeV) of the $\Xi_{bc}$ baryons, corresponding to the Table XII. The horizontal dashed line shows the $\Lambda_{b}D$ threshold.}
\label{fig:Xibc}
\end{figure}

\setlength{\abovecaptionskip}{-0.4cm}
\begin{figure}[!ht]
\begin{center}
\includegraphics[width=0.6\textwidth]{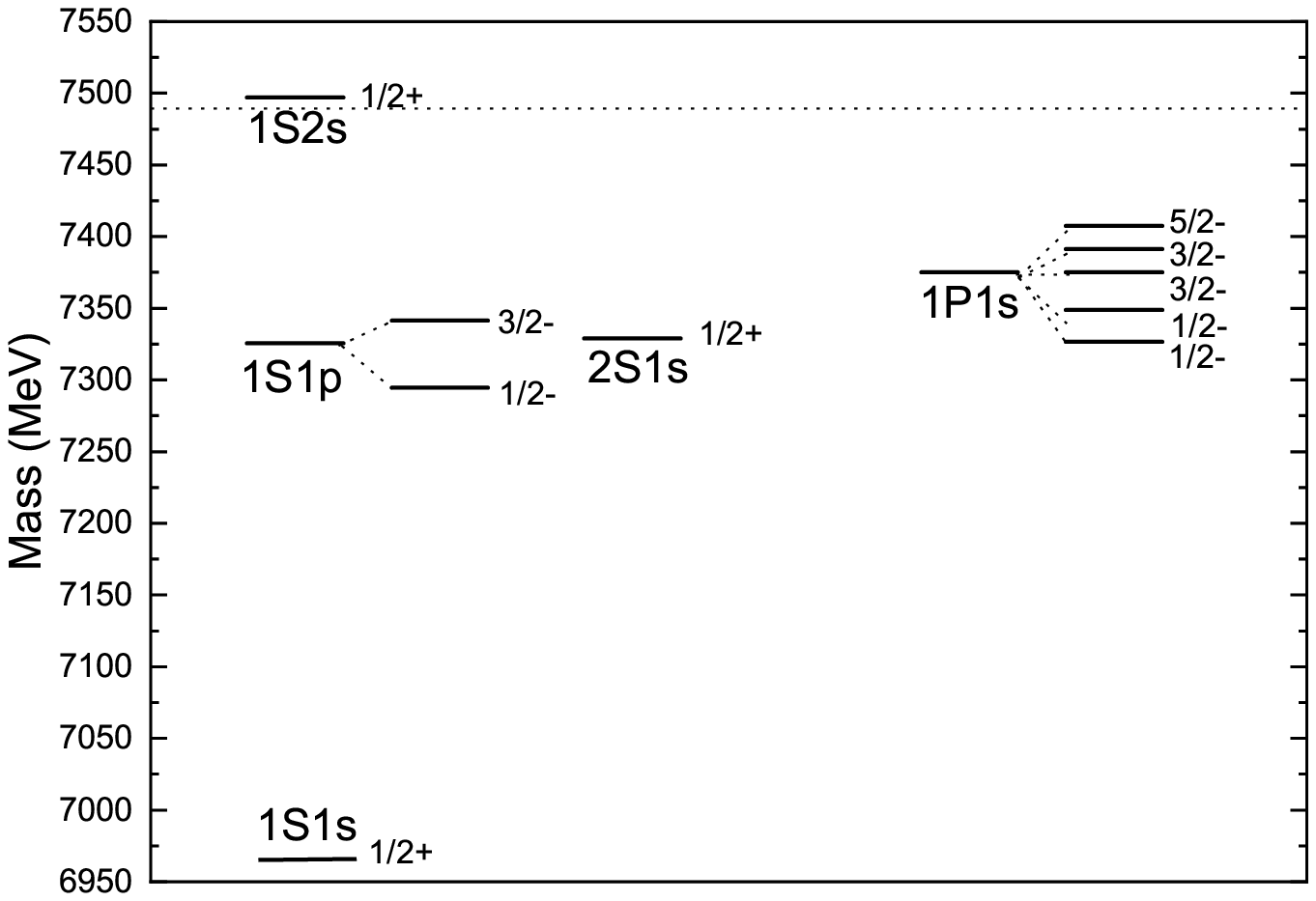}
\end{center}
\caption{Mass spectrum(in GeV) of the $\Xi_{bc}^{\prime}$ baryons, corresponding to the Table XVI. The horizontal dashed line shows the $\Lambda_{b}D$ threshold.}
\label{fig:Xibcp}
\end{figure}

\setlength{\abovecaptionskip}{-0.4cm}
\begin{figure}[!ht]
\begin{center}
\includegraphics[width=0.6\textwidth]{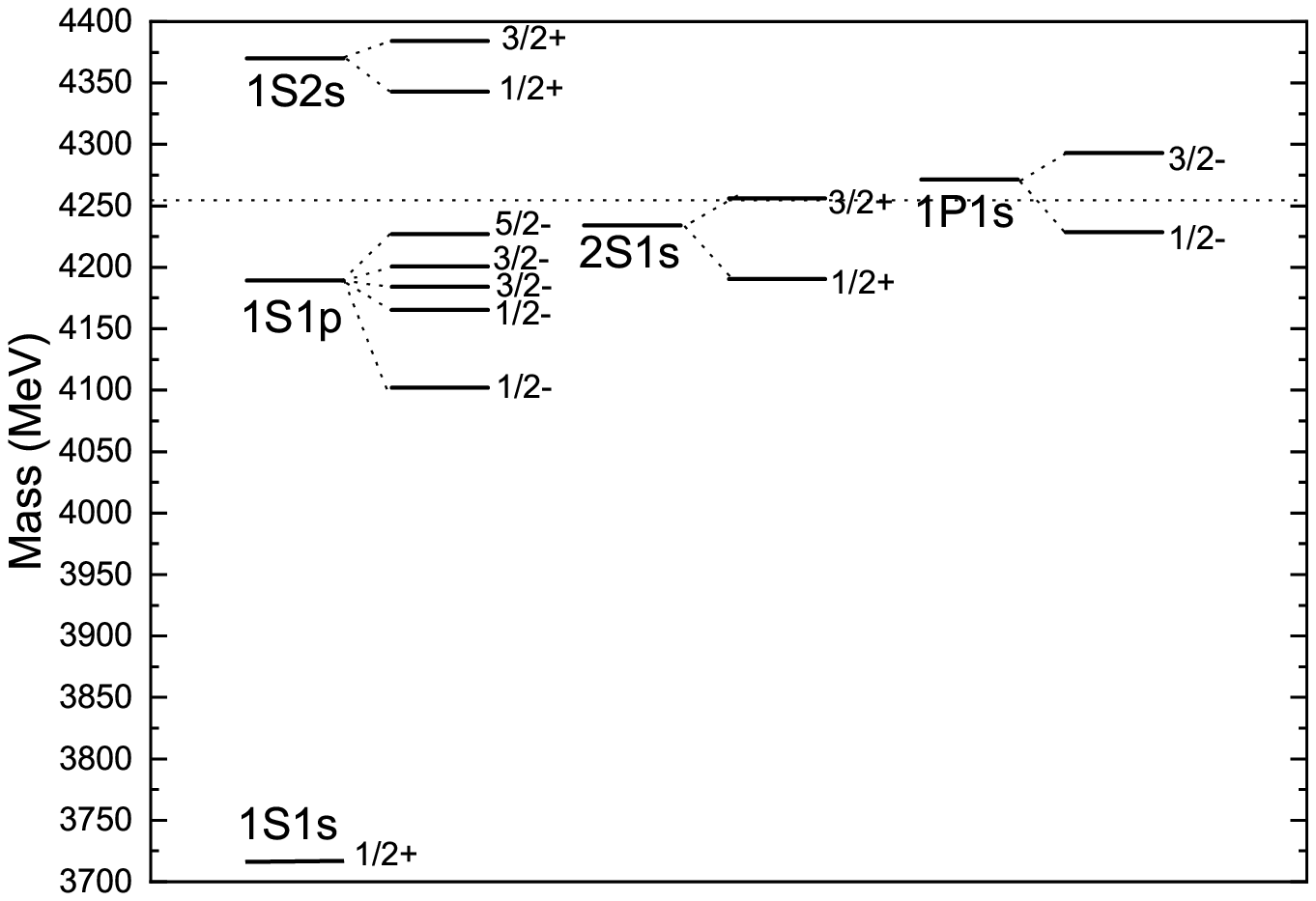}
\end{center}
\caption{Mass spectrum(in MeV) of the $\Omega_{cc}$ baryons, corresponding to the Table XIII. The horizontal dashed line shows the $\Lambda_{c}D_{s}$ threshold.}
\label{fig:Omegacc}
\end{figure}
\setlength{\abovecaptionskip}{-0.6cm}
\begin{figure}[!ht]
\begin{center}
\includegraphics[width=0.6\textwidth]{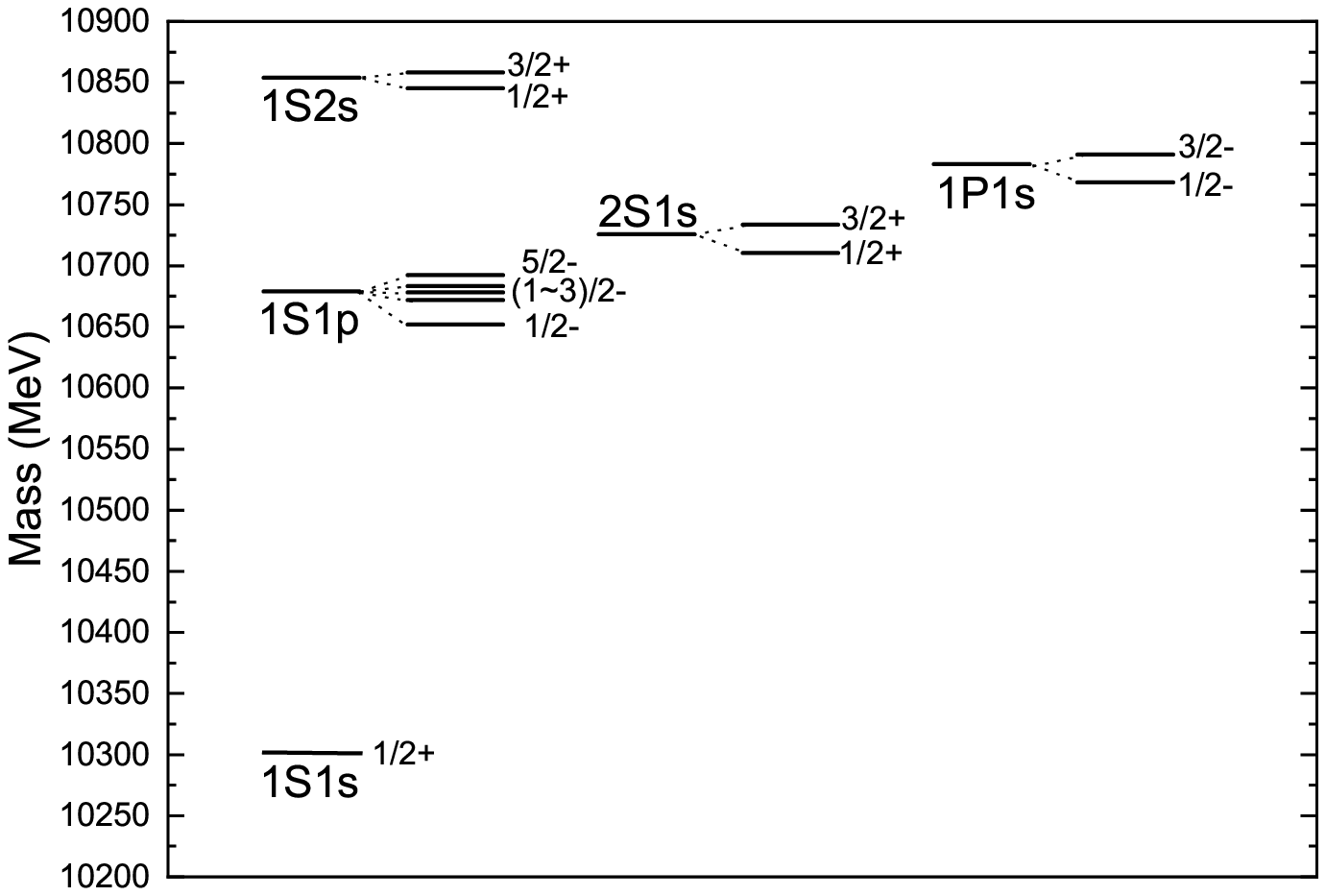}
\end{center}
\caption{Mass spectrum(in MeV) of the $\Omega_{bb}$ baryons, corresponding to the Table XIV.}
\label{fig:Omegabb}
\end{figure}
\setlength{\abovecaptionskip}{-0.4cm}
\begin{figure}[!ht]
\begin{center}
\includegraphics[width=0.6\textwidth]{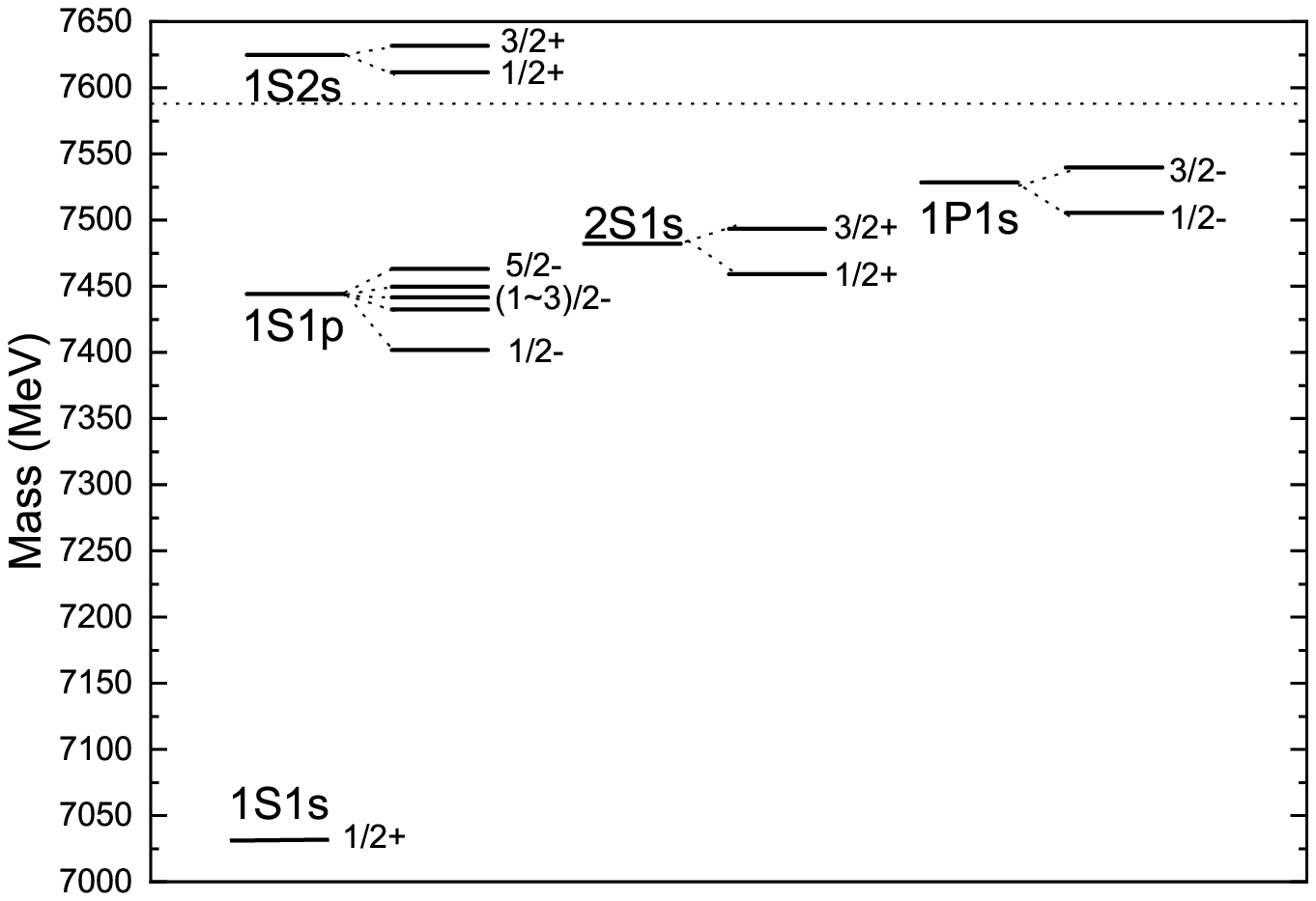}
\end{center}
\caption{Mass spectrum(in MeV) of the $\Omega_{bc}$ baryons, corresponding to the Table XV. The horizontal dashed line shows the $\Lambda_{b}D_{s}$ threshold.}
\label{fig:Omegabc}
\end{figure}
\setlength{\abovecaptionskip}{-0.6cm}
\begin{figure}
\begin{center}
\includegraphics[width=0.6\textwidth]{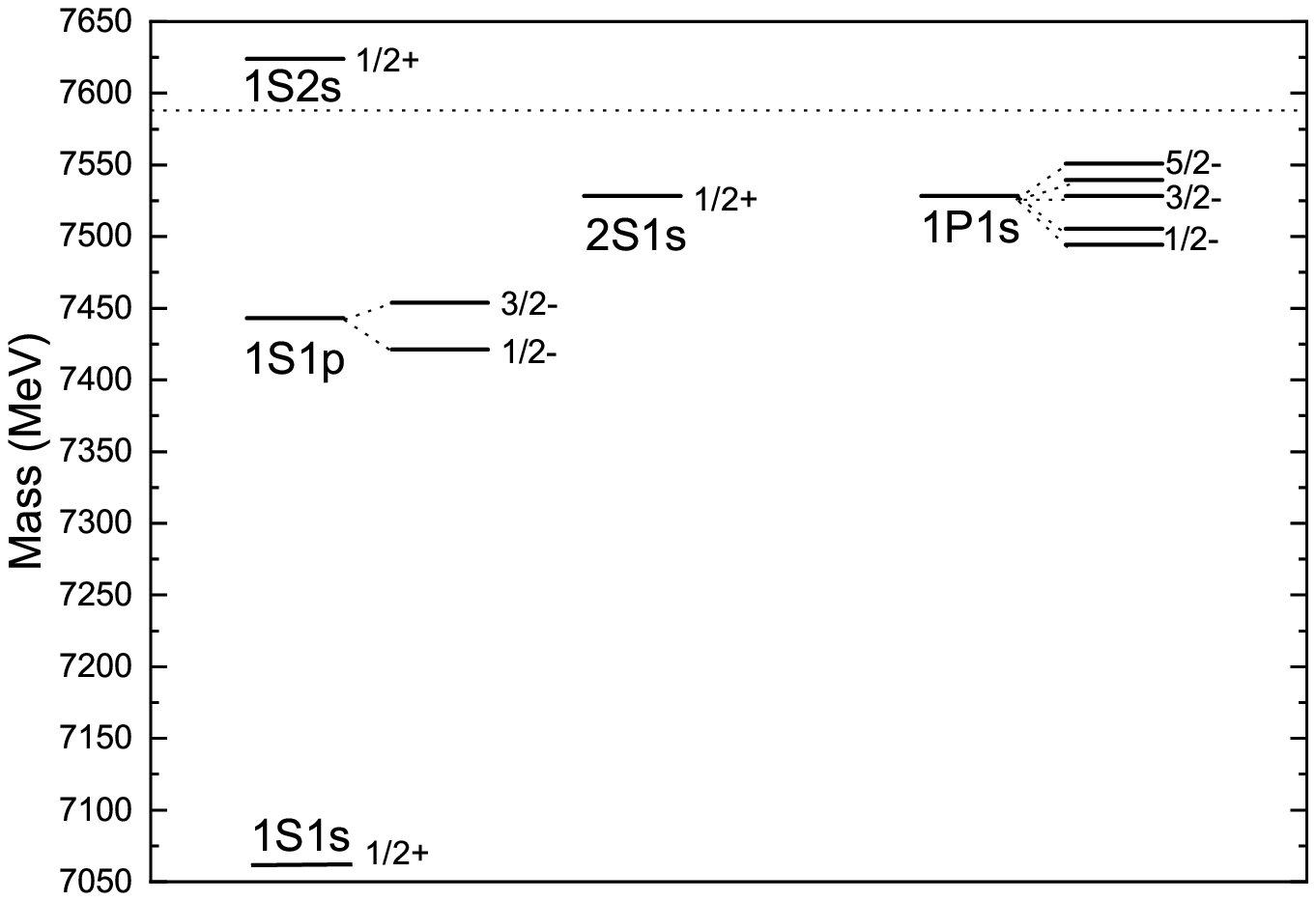}
\end{center}
\caption{Mass spectrum(in MeV) of $\Omega_{bc}^{\prime}$ baryons, corresponding to the Table XVI. The horizontal dashed line shows the $\Lambda_{b}D_{s}$ threshold.}
\label{fig:Omegabcp}
\end{figure}
\vspace{0.2cm}

\section{Summary and remarks}

In this work, we combine methods of Regge trajectory for spin independent mass and mass scaling for the spin-dependent interaction, which is based on chromodynamics similarity among heavy hadrons, to explore excited doubly heavy baryons $\Xi_{QQ^{\prime}}$ and $\Omega_{QQ^{\prime}}$ in the picture of heavy-diquark-light-quark. The low-lying masses of
the excited doubly-heavy baryons are computed up to 2S and 1P waves of the light quark and heavy diquark internally, and compared to other calculations. Two Regge trajectories, one linear and the other nonlinear and endowed with an entra binding of heavy quark pair, are employed to describe the respective excitations of the light quark and diquark, which are derived semiclassically from the QCD string model and tested successfully against the observed heavy baryons and mesons. The heavy-pair binding are extracted from the excitation energies of heavy diquarks phenomenologically and incorporated into Regge relation, by which effective masses of excited diquarks are estimated and mass scaling is constructed.

Our mass analysis suggests (see FIGs. 1-8) that the mean mass-level spacings $\Delta\bar{M}$ of heavy diquark excitations are generally narrower than that of light-quark excitations. This is due to the scale separation of excitations stemming from the adiabatic expansion between the heavy and light-quark dynamics: $\Delta\bar{M}\sim 1/\sqrt{m_{q}} $ is larger than $\Delta\bar{M}\sim 1/\sqrt{\mathrm{Mass}(QQ)}$. Meanwhile, it seems (see FIGs. 1-8) that the mass splitting of spin multiplets of the excited DH baryons are wider for the diquark excitions than for the excitation of light quarks, which remain to be explained in the further explorations. We hope the oncoming experiments like LHCb and Belle to examine our predictions for the excited DH resonances addressed.

The binding function Eq. (\ref{dBNL}), inspired by atomic spectra, mimic the nonrelativistic spectra of a system of heavy quark and antiquark bounded by the Coulomb-like force of single gluon exchange. In a purely Coulombic potential $V(r)=-(4/3)\alpha_{s}/r$, the energy levels  $E_{n}=-[(4/3)\alpha_{s}]^{2}\mu/(2n^{2})$ ($n=n_{r}+L+1$) depends linearly on the reduced mass $\mu$ of system. In a quarkonia or the heavy meson $B_{c}$, Eq. (\ref{dBNL}) holds qualitatively as $B_{0}$ may also depend weakly on the quantum numbers ($N,L$) and upon the effective mass of the quarks\cite{DeRuju:D75,KR:D18,ZXJ:D21}. So, the prediction in Eq.(\ref{Bbcb}) for the 1P-wave binding $B(b\bar{c})$ and ensuing 1P-wave mass calculations of the baryon $\Xi_{bc}$ and $\Omega_{bc}$ are of approximated(uncertain within $10$ MeV roughly).

Low-energy QCD interaction is known to be involved in many aspects,
especially in excited hadrons, and there exist various approaches exploring the excited doubly heavy baryons \cite{Gershtein:2000nx,EFG:D02,Lu:2017meb, Giannuzzi:2009gh,Yoshida:2015tia,Shah:2017liu,Eakins:2012jk,Roberts:2007ni}. The relations of Regge trajectory employed in this work are established phenomenologically and rooted in the underlying interaction of QCD. Some remarks and discussions are in order:

(i) For the DH baryons with 1S-wave diquark, the Regge trajectory stems from
excitation of the light quark $q$ which moves relativistically and is away
from the heavy mass-center of diquark most of time. The short-distance
interaction and binding energy between the heavy diquark and the light quark $q$ is small and ignorable, especially for excited states. In this case, the linear Regge relation applies as in Eq. (\ref{regge0}), without including binding term.

(ii) When diquark of DH baryons excited, an improved(nonlinear) Regge
relation entails a correction due to extra heavy-pair binding, as in Eqs. (\ref{Mbnl}) and (\ref{MQQb}), since the heavy-pairs in hadrons are deep in attractive Coulomb-like potential. This is supported by the energy drop(roughly half or 1/3) of the extracted binding of the 2S and 1P wave baryons relative to the ground state, as indicated by Table IX phenomenologically.

(iii) While the mass predictions for excited DH baryons in this work is by no
means rigorous, our estimation of the baryon spin-multiplet splitting is
general in that mass scaling employed is based on chromodynamics
similarity between heavy baryons and heavy mesons. The parameterized
spin-dependent interactions Eq. (\ref{spin-d}) and Eq. (\ref{HLss}) is built generally according to Lorentz structure and tensor nature of interaction\cite{Landau,MNS}.

Producing and measuring the DH baryons in the $e^{+}e^{-}$, $p\bar{p}$, or $%
pp$ collisions require simultaneous production of two heavy quark-antiquark
pairs and subsequent searches among the final states to which DH baryons
decay. A heavy quark $Q$ from one pair needs to coalesce
with a heavy quark $Q^{\prime }$ from the other pair, forming together a
color antitriplet heavy diquark $QQ^{\prime }$. The heavy diquark $%
QQ^{\prime }$ then needs to pick up a light quark $q$ to finally hadronize
as a DH baryon $QQ^{\prime }q$. To search experimental signals of the DH
baryons, it is useful to check the baryon's decay modes which are easier to
detect. One way is to exmaine the strong decay processes like $\Xi
_{QQ^{\prime }}\rightarrow \Xi _{Q}\pi $ and $\Xi _{QQ^{\prime }}\rightarrow
\Xi _{Q}\rho $($\rho \rightarrow \pi \pi $), which are explored in \cite%
{KL2002,GL19}. The other way is to check the decay modes with two-lepton in
their final states, e.g., the semi-leptonic decays. We list the estimated
lifetimes (except for $\Xi _{cc}^{++}$ and $\Xi _{cc}^{+}$ for which we use
the experimental values) of the DH baryons and some notable branching
fractions ($Br$, all from most recent computation \cite{GL19}) in the
channel of semi-leptonic decay:

$\Xi _{cc}^{++}=ccu$: $\tau =$ $256\pm 27\,$fs(\cite{ZylaPDG:D20}), $Br(\Xi
_{cc}^{++}\rightarrow \Xi _{c}^{+}l\nu _{l})=4.99\%$, $Br(\Xi
_{cc}^{++}\rightarrow \Xi _{c}^{\prime +}l\nu _{l})=5.98\%$;

$\Xi _{cc}^{+}=ccd$: $\tau <33$ fs(\cite{ZylaPDG:D20}), $Br(\Xi _{cc}^{+}\rightarrow \Xi
_{c}^{0}l\nu _{l})=1.65\%$, $Br(\Xi _{cc}^{+}\rightarrow \Xi _{c}^{\prime
0}l\nu _{l})=1.98\%$;

$\Xi _{bc}^{+}=bcu$: $\tau =244$ fs [4], $Br(\Xi _{bc}^{+}\rightarrow \Xi
_{b}^{0}l\nu _{l})=2.3\%$, $Br(\Xi _{bc}^{+}\rightarrow \Xi _{cc}^{++}l\nu
_{l})=1.58\%$;

$\Xi _{bc}^{0}=bcu$: $\tau =93$ fs [4], $Br(\Xi _{bc}^{0}\rightarrow \Xi
_{b}^{-}l\nu _{l})=0.868\%$, $Br(\Xi _{bc}^{0}\rightarrow \Xi _{bcc}^{+}l\nu
_{l})=0.603\%$;

$\Xi _{bb}^{0}=bbu$: $\tau =370$ fs [4], $Br(\Xi _{bb}^{0}\rightarrow \Xi
_{bc}^{+}l\nu _{l})=2.59\%$, $Br(\Xi _{bb}^{0}\rightarrow \Xi _{bc}^{\prime
+}l\nu _{l})=1.15\%$;

$\Xi _{bb}^{-}=bbd$: $\tau =370$ fs [4], $Br(\Xi _{bb}^{-}\rightarrow \Xi
_{bc}^{0}l\nu _{l})=2.62\%$;

$\Omega _{bc}^{0}=bcs$: $\tau =220$ fs \cite{KL2002}, $Br(\Omega
_{bc}^{-}\rightarrow \Omega _{b}^{-}l\nu _{l})=6.03\%$;

$\Omega _{bb}^{-}=bbs$: $\tau =800$ fs \cite{KL2002}, $Br(\Omega
_{bb}^{-}\rightarrow \Omega _{bc}^{0}l\nu _{l})=4.81\%$.

The more details of branching fractions in other channels can be found in
Refs. \cite{GL19,WYZ17} and reference therein. We hope the upcoming
experiments(and data analysis) at LHC, Belle II and CEPC can test our mass
computation of the DH baryons in this work.

\textbf{ACKNOWLEDGMENTS}

Y. S thanks W-N Liu for useful discussions. D. J thanks Xiang Liu and Xiong-Fei Wang for useful discussions. This work is supported by the National Natural Science Foundation of China under Grant No. 12165017.

\section*{\textbf{Appendix A} }

\setcounter{equation}{0}
\renewcommand{\theequation}{A\arabic{equation}}
For the orbitally excitations of a DH baryon $q(QQ^{\prime})$ with S-wave
diquark $QQ^{\prime}$, the classical energy and orbital angular momentum for
the rotating QCD string read
\begin{equation}
E=\sum_{i=QQ^{\prime},q^{\prime}}\left[  \frac{m_{\text{bare}i}}%
{\sqrt{1-\left(  \omega r_{i}\right)  ^{2}}}+\frac{a}{\omega}\int_{0}^{\omega
r_{i}}\frac{du}{\sqrt{1-u^{2}}}\right]  , \label{E}%
\end{equation}%
\begin{equation}
l=\sum_{i=QQ^{\prime},q^{\prime}}\left[  \frac{m_{i}\omega r_{i}^{2}}%
{\sqrt{1-\left(  \omega r_{i}\right)  ^{2}}}+\frac{a}{\omega^{2}}\int
_{0}^{\omega r_{i}}\frac{u^{2}du}{\sqrt{1-u^{2}}}\right]  , \label{L}%
\end{equation}
where $m_{\text{bare}QQ^{\prime}}$ and $m_{\text{bare}q}$ are the respective
bare masses of the heavy diquarks $QQ^{\prime}$ and light quark $q$, and
$\omega r_{i}=v_{i}$ are the velocity of the string end tied to the quark
$i=QQ^{\prime},q$, $a$ stands for the tension of the QCD string. We define the
effective (dynamical) masses of the heavy diquarks and the light quark in the
CM frame of the baryon by
\begin{equation}
M_{QQ^{\prime}}=\frac{m_{\text{bare}QQ^{\prime}}}{\sqrt{1-v_{QQ^{\prime}}^{2}%
}},m_{q}=\frac{m_{\text{bare}q}}{\sqrt{1-v_{q}^{2}}}, \label{effect}%
\end{equation}
Integrating Eq. (\ref{E}) and Eq. (\ref{L}) gives
\begin{equation}
E=M_{QQ^{\prime}}+m_{q}+\frac{a}{\omega}\left[  \arcsin\left(  v_{QQ^{\prime}%
}\right)  +\arcsin\left(  v_{q}\right)  \right]  , \label{E1}%
\end{equation}%
\begin{equation}
l=\frac{1}{\omega}\left(  M_{QQ}v_{QQ^{\prime}}^{2}+m_{q}v_{q}^{2}\right)
+\sum_{i=QQ^{\prime},q}\left[  \arcsin\left(  v_{i}\right)  -v_{i}%
\sqrt{1-v_{i}^{2}}\right]  , \label{L1}%
\end{equation}

The boundary condition of string at ends linked to heavy quark gives
\begin{equation}
\frac{a}{\omega}=\frac{(m_{\text{bare}QQ^{\prime}})v_{QQ^{\prime}}%
}{1-v_{QQ^{\prime}}^{2}}=\frac{M_{QQ}v_{QQ^{\prime}}}{\sqrt{1-v_{QQ^{\prime}%
}^{2}}}. \label{boundary}%
\end{equation}
As the diquark $QQ^{\prime}$ is very heavy and moves nonrelativistically in
hadrons, $v_{QQ}$ is small in the limit of heavy quark($m_{\text{bare}%
QQ^{\prime}}\rightarrow0$). A series expansion of Eq. (\ref{boundary}) upon
$v_{QQ}$ gives
\begin{equation}
\frac{a}{\omega}\simeq M_{QQ}v_{QQ^{\prime}}+\frac{1}{2}M_{QQ}v_{QQ^{\prime}%
}^{3}=P_{QQ^{\prime}}+\frac{P_{QQ^{\prime}}^{3}}{2M_{QQ^{\prime}}^{2}},
\label{bb}%
\end{equation}
From Eq. (\ref{effect}) one has $v_{q}=\sqrt{1-\left(  m_{\text{bare}q}%
/m_{q}\right)  ^{2}}$.

Assuming $q$ to move relativistically($v_{q}\rightarrow1 $),
or, $m_{\text{bare}q}/m_{q}\ll1$, one finds
\begin{equation}
\arcsin\left(  v_{q}\right)  =\arcsin\left(  \sqrt{1-\left(  \frac
{m_{\text{bare}q}}{m_{q}}\right)  ^{2}}\right)  \simeq\frac{\pi}{2}%
-\frac{m_{\text{bare}q}}{m_{q}},\label{d}%
\end{equation}%
\begin{equation}
\arcsin\left(  v_{QQ^{\prime}}\right)  =v_{QQ^{\prime}}+\frac{1}%
{6}v_{QQ^{\prime}}^{3}+\mathcal{O}\left(  v_{QQ^{\prime}}^{5}\right)
,\label{e}%
\end{equation}%
\begin{equation}
v_{QQ^{\prime}}\sqrt{1-v_{QQ^{\prime}}^{2}}=v_{QQ^{\prime}}-\frac{1}%
{2}v_{QQ^{\prime}}^{3}+\mathcal{O}\left(  v_{QQ^{\prime}}^{5}\right)
,\label{f}%
\end{equation}
Substitute the above relations into Eqs. (\ref{E1}) and (\ref{L1}) yields%
\begin{equation}
E\simeq M_{QQ^{\prime}}+m_{q}+\frac{a}{\omega}\left[  \frac{\pi}{2}%
-\frac{m_{\text{bare}q}}{m_{q}}+v_{QQ^{\prime}}+\frac{1}{6}v_{QQ^{\prime}}%
^{3}\right]  ,\label{emq1}%
\end{equation}%
\begin{equation}
\omega l\simeq M_{QQ^{\prime}}v_{QQ^{\prime}}^{2}+m_{q}+\frac{a}{\omega
}\left[  \frac{\pi}{4}-\frac{m_{\text{bare}q}}{m_{q}}\right]  +\frac{\alpha
}{3\omega}v_{QQ^{\prime}}^{3}.\label{wl}%
\end{equation}

Using Eq. (\ref{bb}) and eliminating $\omega$, Eqs. (\ref{emq1}) and
(\ref{wl}) combines to give, when ignoring the tiny term $m_{\text{bare}%
q}/m_{q}$,
\begin{equation}
\left(  E-M_{QQ^{\prime}}\right)  ^{2}=\pi al+\left(  m_{q}+\frac
{P_{QQ^{\prime}}^{2}}{M_{QQ^{\prime}}}\right)  ^{2}-2m_{\text{bare}%
q}P_{QQ^{\prime}}. \label{em}%
\end{equation}
where $P_{QQ^{\prime}}\equiv M_{QQ^{\prime}}v_{QQ^{\prime}}\simeq
M_{QQ^{\prime}}\left(  1-m_{\text{bare}QQ^{\prime}}^{2}/M_{QQ^{\prime}}%
^{2}\right)  ^{1/2}$. Taking the small bare-mass limit($m_{\text{bare}%
q}\rightarrow0$), Eq. (\ref{em}) leads to Eq. (\ref{regge0}), where
$P_{QQ^{\prime}}^{2}/M_{QQ^{\prime}}=M_{QQ^{\prime}}-m_{\text{bare}QQ^{\prime
}}^{2}/M_{QQ^{\prime}}$.

\section*{\textbf{Appendix B} }
\setcounter{equation}{0}
\renewcommand{\theequation}{B\arabic{equation}}

(1)Quantization condition for heavy quarkonia $Q\bar{Q}$

Consider a heavy quarkonia $Q\bar{Q}$ (at a distance of $r$) in a linear
confining potential $T|r|$ for which the system Hamiltonian is $H^{Q\bar{Q}%
}=2\sqrt{\mathbf{p}^{2}+M_{Q}^{2}}+T|r|$. Here, we have ignored the
short-distance interaction. In the frame of center-of-mass, the heavy quark
$Q$ moves equivalently in a confining potential $Tx$, with $x=r/2$, and then
the Hamiltonian for $Q$ as a half system of the quarkonia $Q\bar{Q}$, becomes%
\begin{equation}
H^{Q}=\frac{1}{2}H^{Q\bar{Q}}=\sqrt{p_{x}^{2}+l_{q}^{2}/x^{2}+M_{Q}^{2}%
}+T|x|.\label{HQ}%
\end{equation}
Using the semiclassical WKB analysis upon Eq. (\ref{HQ}) for the radial
excitations($l_{q}=0$), one has a WKB quantization
condition\cite{SonnenW:jh2014}\footnote{It is written as $\int_{x_{-}}^{x_{+}%
}p(x)dx=\pi N$ in the cited literature.},%
\begin{equation}
2\int_{x_{\_}}^{x_{+}}p_{x}(x)dx=2\pi(N+c_{0}),\label{qc0}%
\end{equation}
with $p_{x}(x)=|\mathbf{p}_{x}|=\sqrt{[\left(  E(Q)-T|x|\right)  ^{2}%
-M_{Q}^{2}]}$, $x_{+}=(E(Q)-M_{Q})/T=-x_{-}$ the classical \textquotedblleft
turning points\textquotedblright, $N$ the radial quantum number, $c_{0}$ a
constant. Here, $E(Q)$ is the semiclassical value of $H^{Q}$, and the factor 2
before the integral in Eq. (\ref{qc0}) arises from underlying spinor nature of
a quark whose wave function returns to original value after double cycle of
journey in the position space\cite{JiaD:E19}. Assuming $Q$ to be in S
wave(moving radially), integration of Eq. (\ref{qc0}) gives
\begin{equation}
\pi(N+c_{0})=\frac{E(Q)^{2}}{T}\left[  \sqrt{1-B^{2}}+B^{2}\ln\left(
\frac{1-\sqrt{1-B^{2}}}{B}\right)  \right]  ,\label{qc1}%
\end{equation}
with $B\equiv M_{Q}/E(Q)$.

Above result is only for the half system. Transforming to the whole system of
quarkonia by mapping $E(Q)\rightarrow E/2$ and $N\rightarrow N/2$%
\cite{SonnenW:jh2014}, Eq. (\ref{qc1}) gives($B\rightarrow\tilde{B}=2M_{Q}%
/E$)
\begin{align}
2\pi T(N+2c_{0})  &  =E^{2}\left[  \sqrt{1-\tilde{B}^{2}}+\tilde{B}^{2}%
\ln\left(  \frac{1-\sqrt{1-\tilde{B}^{2}}}{\tilde{B}}\right)  \right]
\nonumber\\
&  \simeq E^{1/2}\left[  \frac{4\sqrt{2}}{3}E^{3/2}\epsilon^{3/2}-\frac
{7\sqrt{2}}{15}E^{3/2}\epsilon^{5/2}+\mathcal{O}(\epsilon^{7/2})\right]
\label{qc2}%
\end{align}
where $\epsilon\equiv1-\tilde{B}=1-2M_{Q}/E$, which is small in heavy quark
limit and $E=2M_{Q}/(1-\epsilon)\simeq2M_{Q}$. Thus, Eq. (\ref{qc2}) leads to,
to the leading order of $\epsilon$,
\[
3\pi T(N+2c_{0})=4\sqrt{M_{Q}}(E-2M_{Q})^{3/2},
\]
which gives Eq. (\ref{RadN}).

(2) Improved(nonlinear) Regge relation

Rewriting in a typical form of standard trajectory, $\left(  \bar
{M}-const.\right) ^{3/2}\sim$ quantum numbers, for the heavy quarkonia system,
the Regge relation Eq. (\ref{M23}) in Ref. \cite{BurnsPP:D10} becomes
\begin{equation}
\left(  \bar{M}-2M_{Q}\right)  ^{3/2}=\frac{3\sqrt{3}}{2\sqrt{M_{Q}}%
}TL.\label{typ}%
\end{equation}
Comparing the radial and angular slopes(linear coefficients in $N$ and $L$) in
RHS of Eq. (\ref{RadN}) and Eq. (\ref{typ}), which is $\pi:\sqrt{12}$, one can
combine two equations into one unified form:
\begin{equation}
\left[  \bar{M}_{N,L}-2M_{Q}\right]  ^{3/2}=\frac{3\sqrt{3}T}{2\sqrt{M_{Q}}%
}\left(  L+\frac{\pi N}{\sqrt{12}}+2c_{0}\right)  .\label{Mcom}%
\end{equation}

In the derivation\cite{BurnsPP:D10} of Eq. (\ref{M23}) and that of Eq.
(\ref{RadN}) shown in Appendix Eq.(\ref{HQ}), the linear confining interaction between $Q$ and $\bar{Q}$ is assumed, with the short-distance force between them ignored. As
the short-distance force is required for low-lying quarkonia system and violates the typical linear trajectory in Eq. (\ref{Mcom}), one way out is to assume that the short-distance attractive force between $Q$ and $\bar{Q}$ in the
color-singlet($1_{c}$) would provide an extra(negative) energy $B(Q\bar
{Q})_{N,L}$ to the spectra of the $Q\bar{Q}$ system and to deduct $B(Q\bar
{Q})_{N,L}$ (named the binding energy) from the hadron mass $\bar{M}_{N,L}$ in
Eq. (\ref{Mcom}) so that the LHS of Eq. (\ref{Mcom}) becomes the remaining
string energy solely: $(\bar{M}_{N,L}-2M_{Q})^{3/2}\rightarrow(\bar{M}%
_{N,L}-B(Q\bar{Q})_{N,L}-2M_{Q})^{3/2}$. In doing so, the arguments by the
classical string picture\cite{BurnsPP:D10} and that in Eq. (1) given above remain
valid and the formula Eq. (\ref{Mcom}) remains intact formally up to a replacement
$\bar{M}_{N,L}\rightarrow\bar{M}_{N,L}-B(Q\bar{Q})_{N,L}$. This gives rise to
Eq. (\ref{MNL}).

The binding energy $B(Q\bar{Q})_{N,L}$(in the excited states $|N,L\rangle$ of
system) depends on the quantum numbers($N,L$) of the system considered
\cite{KR:D18} and thereby violates the linearity of the Regge relations Eq. (\ref{typ}) and Eq. (\ref{RadN}). Normally, such a term is negative since when two
quarks($Q$ and $\bar{Q}$) are heavy enough to stay close each other, they both
experiences an attractive Coulomb-like force of single gluon exchange, as the
observed heavy-quarkonia spectra\cite{ZylaPDG:D20} of the cascade type indicated.

\section*{\textbf{Appendix C}}

\setcounter{equation}{0}
\renewcommand{\theequation}{C\arabic{equation}}
In the QCD string picture, one can view a DH baryons $qQQ^{\prime}$, with
excited diquark $QQ^{\prime}$, as a string system $[Q-Q]q$, consisting of a
(heavy) subsystem of massive string $[Q-Q]$ (each $Q$ at one of ends) and a
light subsystem of a light-quark $q$ and the string connected to it.

In the semiclassical approximation, one can assume the light subsystem of $q$
and attached string to it to be in stationary state while the heavy subsystem
$[Q-Q]$ is excited to the excited state(denoted by $|N,L\rangle$, say). As
such, the excitation of the heavy subsystem $[Q-Q]$ in color antitriplet($\bar
{3}_{c}$) with string tension $T_{QQ}$ resembles the excitation of a heavy
quarkonia $Q\bar{Q}$ in color singlet($1_{c}$) with string tension $T$ up to a
color(strength) factor of string interaction which is taken to be half
commonly. Based on this similarity, one can write the excitation energy of the
heavy system, by analogy with Eq. (\ref{Mbnl}),
\begin{equation}
\bar{M}(QQ)_{N,L}-2M_{Q}=\Delta B(QQ)_{N,L}+3\frac{\left[  T_{QQ}\left(
(L+\pi N/\sqrt{12})+2c_{0}\right)  \right]  ^{2/3}}{[4M_{Q}]^{1/3}}+c_{1},
\label{MBQQ}%
\end{equation}
in which the first term $\Delta B(QQ)_{N,L}=B(QQ)_{N,L}-B(QQ)_{0,0}$ in the
RSH accounts for the short-distance contribution due to heavy-quark binding
and the second term for the excited string energy of the heavy subsystem in
the $|N,L\rangle$ state, $c_{0}$ is given by Eq. (\ref{Tc0M}). We add an
addictive constant $c_{1}$ since $\bar{M}(QQ)$ is defined up to the ground
state of the whole DH system containing $QQ$.

Extending Eq. (\ref{MBQQ}) to the diquark case of $QQ=bc$, one gets Eq.
(\ref{MQQb}) generally, by heavy quark symmetry.

\section*{\textbf{Appendix D}}

\setcounter{equation}{0}
\renewcommand{\theequation}{D\arabic{equation}}
Given the eigenvalues $j=1/2$ and $3/2$, one solves the bases(eigenfunctions)
$\left\vert S_{QQ^{\prime}3},S_{q3},l_{3}\right\rangle $ of the $\mathbf{l}%
\cdot\mathbf{S}_{q}$ in the $LS$ coupling. The mass formula $\Delta M=\langle
H^{SD}\rangle$ for a DH baryon $qQQ^{\prime}$ with S-wave diquark $QQ^{\prime
}$ can be obtained by diagonalizing the dominate interaction $a_{1}%
\mathbf{l}\cdot\mathbf{S}_{q}$ and adding the diagonal elements of other
perturbative spin-interactions in Eq.~(\ref{spin-d}). This can be done by
evaluating the matrix elements of $H^{SD}$ in the $LS$ coupling and then
changing the bases $\left\vert S_{QQ^{\prime}3},S_{q3},l_{3}\right\rangle $ to
the new bases $\left\vert J,j\right\rangle $ in the $jj$ coupling to find the
mass formula $\Delta M=\langle H^{SD}\rangle$.

For each interaction terms of Eq. (\ref{spin-d}), one can evaluate its matrix
elements by explicit construction of states with a given $J_{3}$ as linear
combinations of the baryon states $\left\vert S_{QQ^{\prime}3},S_{q3}%
,l_{3}\right\rangle $ in the $LS$ coupling where $S_{q3}+S_{QQ^{\prime}%
3}+l_{3}=J_{3}$. Due to the rotation invariance of the matrix elements, it
suffices to use a single $J_{3}$ for each term. Then, one can use
\begin{equation}
\mathbf{l}\cdot\mathbf{S}_{i}=\frac{1}{2}\left[  l_{+}S_{i-}+l_{-}%
S_{i+}\right]  +l_{3}S_{i3}, \label{LS}%
\end{equation}
to find their elements by applying $\mathbf{l}\cdot\mathbf{S}_{i}$
($i=q,QQ^{\prime}$) on the the third components of angular momenta. For
projected states of baryon with given $J_{3}$, they can be expressed, in the
$LS$ coupling, as
\begin{align}
&  \left\vert {\ }^{2}P_{J=1/2},J_{3}=\frac{1}{2}\right\rangle =\frac{\sqrt
{2}}{3}\left\vert 1,-\frac{1}{2},0\right\rangle -\frac{1}{3}\left\vert
0,\frac{1}{2},0\right\rangle \nonumber\\
&  \quad\quad\quad\quad\quad\quad\quad-\frac{\sqrt{2}}{3}\left\vert
0,-\frac{1}{2},1\right\rangle +\frac{2}{3}\left\vert -1,\frac{1}%
{2},1\right\rangle , \label{2P}%
\end{align}%
\begin{align}
&  \left\vert {\ }^{4}P_{J=1/2},J_{3}=\frac{1}{2}\right\rangle =\frac{1}%
{\sqrt{2}}\left\vert 1,\frac{1}{2},-1\right\rangle -\frac{1}{3}\left\vert
1,-\frac{1}{2},0\right\rangle \nonumber\\
&  \quad\quad\quad-\frac{\sqrt{2}}{3}\left\vert 0,\frac{1}{2},0\right\rangle
+\frac{1}{3}\left\vert 0,-\frac{1}{2},1\right\rangle +\frac{1}{3\sqrt{2}%
}\left\vert -1,\frac{1}{2},1\right\rangle , \label{4P}%
\end{align}%
\[
\left\vert {\ }^{2}P_{J=3/2},J_{3}=\frac{3}{2}\right\rangle =\sqrt{\frac{2}%
{3}}\left\vert 1,-\frac{1}{2},1\right\rangle -\sqrt{\frac{1}{3}}\left\vert
0,\frac{1}{2},1\right\rangle ,
\]%
\begin{align}
&  \left\vert {\ }^{4}P_{J=3/2},J_{3}=\frac{3}{2}\right\rangle =\sqrt{\frac
{3}{5}}\left\vert 1,\frac{1}{2},0\right\rangle -\sqrt{\frac{2}{15}}\left\vert
1,-\frac{1}{2},1\right\rangle \nonumber\label{4}\\
&  \quad\quad\quad\quad\quad\quad\quad-\frac{2}{\sqrt{15}}\left\vert
0,\frac{1}{2},1\right\rangle ,
\end{align}%
\[
\left\vert {\ }^{4}P_{J=5/2},\quad J_{3}=\frac{5}{2}\right\rangle =\left\vert
1,\frac{1}{2},1\right\rangle .
\]
and use them to compute the matrix elements of $\mathbf{l}\cdot\mathbf{S}_{i}%
$, $\mathbf{S}_{QQ^{\prime}}\cdot\mathbf{S}_{q}$ and $S_{12}/2$ in the basis
[$^{2}P_{J},^{4}P_{J}$]. With some algebra, one can find
\begin{equation}
\left\langle \mathbf{l}\cdot\mathbf{S}_{QQ^{\prime}}\right\rangle
_{J=1/2}=\left[
\begin{array}
[c]{cc}%
-\frac{4}{3} & -\frac{\sqrt{2}}{3}\\
-\frac{\sqrt{2}}{3} & -\frac{5}{3}%
\end{array}
\right]  ,\left\langle \mathbf{l}\cdot\mathbf{S}_{q}\right\rangle
_{J=1/2}=\left[
\begin{array}
[c]{cc}%
\frac{1}{3} & \frac{\sqrt{2}}{3}\\
\frac{\sqrt{2}}{3} & -\frac{5}{6}%
\end{array}
\right]  , \label{6}%
\end{equation}%
\begin{equation}
\left\langle \mathbf{l}\cdot\mathbf{S}_{QQ^{\prime}}\right\rangle
_{J=3/2}=\left[
\begin{array}
[c]{cc}%
\frac{2}{3} & -\frac{\sqrt{5}}{3}\\
-\frac{\sqrt{5}}{3} & -\frac{2}{3}%
\end{array}
\right]  ,\left\langle \mathbf{l}\cdot\mathbf{S}_{q}\right\rangle
_{J=3/2}=\left[
\begin{array}
[c]{cc}%
-\frac{1}{6} & \frac{\sqrt{5}}{3}\\
\frac{\sqrt{5}}{3} & -\frac{1}{3}%
\end{array}
\right]  , \label{7}%
\end{equation}%
\begin{equation}
\left\langle \mathbf{S}_{QQ^{\prime}}\cdot\mathbf{S}_{q}\right\rangle
_{J=1/2}=\left[
\begin{array}
[c]{cc}%
-1 & 0\\
0 & \frac{1}{2}%
\end{array}
\right]  ,\quad\left\langle \mathbf{S}_{QQ^{\prime}}\cdot\mathbf{S}%
_{q}\right\rangle _{J=3/2}=\left[
\begin{array}
[c]{cc}%
-1 & 0\\
0 & \frac{1}{2}%
\end{array}
\right]  , \label{8}%
\end{equation}%
\begin{equation}
\left\langle S_{12} \right\rangle _{J=1/2}=\left[
\begin{array}
[c]{cc}%
0 & \frac{\sqrt{2}}{2}\\
\frac{\sqrt{2}}{2} & -1
\end{array}
\right]  ,\quad\left\langle S_{12} \right\rangle _{J=3/2}=\left[
\begin{array}
[c]{cc}%
0 & -\frac{\sqrt{5}}{10}\\
-\frac{\sqrt{5}}{10} & \frac{4}{5}%
\end{array}
\right]  , \label{9}%
\end{equation}
In the subspace of $J=5/2$, one finds
\begin{equation}
\left\langle \mathbf{l}\cdot\mathbf{S}_{QQ^{\prime}}\right\rangle
_{J=5/2}=1,\quad\left\langle \mathbf{l}\cdot\mathbf{S}_{q}\right\rangle
_{J=5/2}=\frac{1}{2}, \label{10}%
\end{equation}%
\begin{equation}
\left\langle \mathbf{S}_{QQ^{\prime}}\cdot\mathbf{S}_{q}\right\rangle
_{J=5/2}=\frac{1}{2},\quad\left\langle S_{12}  \right\rangle
_{J=5/2}=-\frac{1}{5}, \label{11}%
\end{equation}

Given the above matrices, one can solve each eigenvalue $\lambda$ of
$\mathbf{l}\cdot\mathbf{S}_{q}$ and the corresponding eigenvectors for a given
$J$, and for that $J$ one can write the baryon states $\left\vert
J,j\right\rangle $ in the $jj$ coupling. They are linear combinations of the
$LS$ bases $\left\vert ^{2S+1}P_{J}\right\rangle $ with definite coefficients
specified by the respective eigenvector solved for $J$:
\begin{equation}
\lambda=-1:\left\vert J=\frac{1}{2},j=\frac{1}{2}\right\rangle =\frac{1}%
{3}\left\vert 1^{2}P_{1/2}\right\rangle -\frac{2\sqrt{2}}{3}\left\vert
1^{4}P_{1/2}\right\rangle ,\label{12}%
\end{equation}%
\begin{equation}
\lambda=+\frac{1}{2}:\left\vert J=\frac{1}{2},j=\frac{3}{2}\right\rangle
=\frac{2\sqrt{2}}{3}\left\vert 1^{2}P_{1/2}\right\rangle +\frac{1}%
{3}\left\vert 1^{4}P_{1/2}\right\rangle ,\label{13}%
\end{equation}%
\begin{equation}
\lambda=-1:\left\vert J=\frac{3}{2},j=\frac{1}{2}\right\rangle =\frac{2}%
{3}\left\vert 1^{2}P_{3/2}\right\rangle -\frac{\sqrt{5}}{3}\left\vert
1^{4}P_{3/2}\right\rangle ,\label{14}%
\end{equation}%
\begin{equation}
\lambda=+\frac{1}{2}:\left\vert J=\frac{3}{2},j=\frac{3}{2}\right\rangle
=\frac{\sqrt{5}}{3}\left\vert 1^{2}P_{3/2}\right\rangle +\frac{2}{3}\left\vert
1^{4}P_{3/2}\right\rangle ,\label{15}%
\end{equation}%
\begin{equation}
\lambda=+\frac{1}{2}:\left\vert J=\frac{5}{2},j=\frac{3}{2}\right\rangle
=\left\vert 1^{4}P_{5/2}\right\rangle ,\label{16}%
\end{equation}
This gives the required baryon states in the heavy diquark limit, by which the
diagonal matrix elements of $\mathbf{l}\cdot\mathbf{S}_{QQ^{\prime}}$,
$S_{12}/2$ and $\mathbf{S}_{QQ^{\prime}}\cdot\mathbf{S}_{q}$ can be obtained.
The detailed results are collected in Table IV.

\end{document}